\documentclass[aps,prl,twocolumn,superscriptaddress,showpacs,preprintnumbers,amsmath,amssymb]{revtex4-1}
\usepackage[utf8]{inputenc}
\usepackage{graphicx}
\usepackage{dcolumn}
\usepackage{bm}
\usepackage{notes2bib}
\usepackage{graphicx}
\usepackage{gensymb}
\usepackage{url}
\usepackage{amsmath}
\usepackage{latexsym}
\usepackage{amssymb}
\usepackage{color}

\usepackage{slashed}
\newcommand{\cost}{{\cos}\theta}
{}

\begin{document}

\title{\quad\\[0.5cm]Observation of Transverse $\Lambda/\bar{\Lambda}$ Hyperon Polarization in $e^+e^-$ Annihilation at Belle}


\date{\today}
\noaffiliation
\affiliation{University of the Basque Country UPV/EHU, 48080 Bilbao}
\affiliation{Beihang University, Beijing 100191}
\affiliation{University of Bonn, 53115 Bonn}
\affiliation{Brookhaven National Laboratory, Upton, New York 11973}
\affiliation{Budker Institute of Nuclear Physics SB RAS, Novosibirsk 630090}
\affiliation{Faculty of Mathematics and Physics, Charles University, 121 16 Prague}
\affiliation{Chonnam National University, Kwangju 660-701}
\affiliation{University of Cincinnati, Cincinnati, Ohio 45221}
\affiliation{Deutsches Elektronen--Synchrotron, 22607 Hamburg}
\affiliation{Duke University, Durham, North Carolina 27708}
\affiliation{University of Florida, Gainesville, Florida 32611}
\affiliation{Key Laboratory of Nuclear Physics and Ion-beam Application (MOE) and Institute of Modern Physics, Fudan University, Shanghai 200443}
\affiliation{Justus-Liebig-Universit\"at Gie\ss{}en, 35392 Gie\ss{}en}
\affiliation{Gifu University, Gifu 501-1193}
\affiliation{II. Physikalisches Institut, Georg-August-Universit\"at G\"ottingen, 37073 G\"ottingen}
\affiliation{SOKENDAI (The Graduate University for Advanced Studies), Hayama 240-0193}
\affiliation{Gyeongsang National University, Chinju 660-701}
\affiliation{Hanyang University, Seoul 133-791}
\affiliation{University of Hawaii, Honolulu, Hawaii 96822}
\affiliation{High Energy Accelerator Research Organization (KEK), Tsukuba 305-0801}
\affiliation{J-PARC Branch, KEK Theory Center, High Energy Accelerator Research Organization (KEK), Tsukuba 305-0801}
\affiliation{Forschungszentrum J\"{u}lich, 52425 J\"{u}lich}
\affiliation{IKERBASQUE, Basque Foundation for Science, 48013 Bilbao}
\affiliation{Indian Institute of Science Education and Research Mohali, SAS Nagar, 140306}
\affiliation{Indian Institute of Technology Bhubaneswar, Satya Nagar 751007}
\affiliation{Indian Institute of Technology Guwahati, Assam 781039}
\affiliation{Indian Institute of Technology Hyderabad, Telangana 502285}
\affiliation{Indian Institute of Technology Madras, Chennai 600036}
\affiliation{Indiana University, Bloomington, Indiana 47408}
\affiliation{Institute of High Energy Physics, Chinese Academy of Sciences, Beijing 100049}
\affiliation{Institute of High Energy Physics, Vienna 1050}
\affiliation{Institute for High Energy Physics, Protvino 142281}
\affiliation{INFN - Sezione di Napoli, 80126 Napoli}
\affiliation{INFN - Sezione di Torino, 10125 Torino}
\affiliation{Advanced Science Research Center, Japan Atomic Energy Agency, Naka 319-1195}
\affiliation{J. Stefan Institute, 1000 Ljubljana}
\affiliation{Kanagawa University, Yokohama 221-8686}
\affiliation{Institut f\"ur Experimentelle Teilchenphysik, Karlsruher Institut f\"ur Technologie, 76131 Karlsruhe}
\affiliation{Kennesaw State University, Kennesaw, Georgia 30144}
\affiliation{King Abdulaziz City for Science and Technology, Riyadh 11442}
\affiliation{Department of Physics, Faculty of Science, King Abdulaziz University, Jeddah 21589}
\affiliation{Korea Institute of Science and Technology Information, Daejeon 305-806}
\affiliation{Korea University, Seoul 136-713}
\affiliation{Kyoto University, Kyoto 606-8502}
\affiliation{Kyungpook National University, Daegu 702-701}
\affiliation{LAL, Univ. Paris-Sud, CNRS/IN2P3, Universit\'{e} Paris-Saclay, Orsay}
\affiliation{\'Ecole Polytechnique F\'ed\'erale de Lausanne (EPFL), Lausanne 1015}
\affiliation{P.N. Lebedev Physical Institute of the Russian Academy of Sciences, Moscow 119991}
\affiliation{Faculty of Mathematics and Physics, University of Ljubljana, 1000 Ljubljana}
\affiliation{Ludwig Maximilians University, 80539 Munich}
\affiliation{Luther College, Decorah, Iowa 52101}
\affiliation{University of Malaya, 50603 Kuala Lumpur}
\affiliation{University of Maribor, 2000 Maribor}
\affiliation{Max-Planck-Institut f\"ur Physik, 80805 M\"unchen}
\affiliation{School of Physics, University of Melbourne, Victoria 3010}
\affiliation{University of Mississippi, University, Mississippi 38677}
\affiliation{University of Miyazaki, Miyazaki 889-2192}
\affiliation{Moscow Physical Engineering Institute, Moscow 115409}
\affiliation{Moscow Institute of Physics and Technology, Moscow Region 141700}
\affiliation{Graduate School of Science, Nagoya University, Nagoya 464-8602}
\affiliation{Kobayashi-Maskawa Institute, Nagoya University, Nagoya 464-8602}
\affiliation{Universit\`{a} di Napoli Federico II, 80055 Napoli}
\affiliation{Nara Women's University, Nara 630-8506}
\affiliation{National Central University, Chung-li 32054}
\affiliation{National United University, Miao Li 36003}
\affiliation{Department of Physics, National Taiwan University, Taipei 10617}
\affiliation{H. Niewodniczanski Institute of Nuclear Physics, Krakow 31-342}
\affiliation{Nippon Dental University, Niigata 951-8580}
\affiliation{Niigata University, Niigata 950-2181}
\affiliation{Novosibirsk State University, Novosibirsk 630090}
\affiliation{Osaka City University, Osaka 558-8585}
\affiliation{Pacific Northwest National Laboratory, Richland, Washington 99352}
\affiliation{Panjab University, Chandigarh 160014}
\affiliation{Peking University, Beijing 100871}
\affiliation{University of Pittsburgh, Pittsburgh, Pennsylvania 15260}
\affiliation{Theoretical Research Division, Nishina Center, RIKEN, Saitama 351-0198}
\affiliation{RIKEN BNL Research Center, Upton, New York 11973}
\affiliation{University of Science and Technology of China, Hefei 230026}
\affiliation{Showa Pharmaceutical University, Tokyo 194-8543}
\affiliation{Soongsil University, Seoul 156-743}
\affiliation{University of South Carolina, Columbia, South Carolina 29208}
\affiliation{Stefan Meyer Institute for Subatomic Physics, Vienna 1090}
\affiliation{Sungkyunkwan University, Suwon 440-746}
\affiliation{School of Physics, University of Sydney, New South Wales 2006}
\affiliation{Department of Physics, Faculty of Science, University of Tabuk, Tabuk 71451}
\affiliation{Tata Institute of Fundamental Research, Mumbai 400005}
\affiliation{Excellence Cluster Universe, Technische Universit\"at M\"unchen, 85748 Garching}
\affiliation{Department of Physics, Technische Universit\"at M\"unchen, 85748 Garching}
\affiliation{Toho University, Funabashi 274-8510}
\affiliation{Department of Physics, Tohoku University, Sendai 980-8578}
\affiliation{Earthquake Research Institute, University of Tokyo, Tokyo 113-0032}
\affiliation{Department of Physics, University of Tokyo, Tokyo 113-0033}
\affiliation{Tokyo Institute of Technology, Tokyo 152-8550}
\affiliation{Tokyo Metropolitan University, Tokyo 192-0397}
\affiliation{Virginia Polytechnic Institute and State University, Blacksburg, Virginia 24061}
\affiliation{Wayne State University, Detroit, Michigan 48202}
\affiliation{Yamagata University, Yamagata 990-8560}
\affiliation{Yonsei University, Seoul 120-749}
  \author{Y.~Guan}\affiliation{Indiana University, Bloomington, Indiana 47408}\affiliation{High Energy Accelerator Research Organization (KEK), Tsukuba 305-0801}\affiliation{Duke University, Durham, North Carolina 27708}
  \author{A.~Vossen}\affiliation{Duke University, Durham, North Carolina 27708}\affiliation{Indiana University, Bloomington, Indiana 47408} 
  \author{I.~Adachi}\affiliation{High Energy Accelerator Research Organization (KEK), Tsukuba 305-0801}\affiliation{SOKENDAI (The Graduate University for Advanced Studies), Hayama 240-0193} 
  \author{K.~Adamczyk}\affiliation{H. Niewodniczanski Institute of Nuclear Physics, Krakow 31-342} 
  \author{J.~K.~Ahn}\affiliation{Korea University, Seoul 136-713} 
  \author{H.~Aihara}\affiliation{Department of Physics, University of Tokyo, Tokyo 113-0033} 
  \author{S.~Al~Said}\affiliation{Department of Physics, Faculty of Science, University of Tabuk, Tabuk 71451}\affiliation{Department of Physics, Faculty of Science, King Abdulaziz University, Jeddah 21589} 
  \author{D.~M.~Asner}\affiliation{Brookhaven National Laboratory, Upton, New York 11973} 
  \author{H.~Atmacan}\affiliation{University of South Carolina, Columbia, South Carolina 29208} 
  \author{V.~Aulchenko}\affiliation{Budker Institute of Nuclear Physics SB RAS, Novosibirsk 630090}\affiliation{Novosibirsk State University, Novosibirsk 630090} 
% ...
\collaboration{The Belle Collaboration}


\begin{abstract}
We report the first observation of the {spontaneous} polarization of $\Lambda$ and $\bar{\Lambda}$ hyperons transverse to the production plane in $e^+e^-$ annihilation, {which is attributed to the effect arising from {a} polarizing fragmentation function.}
For inclusive $\Lambda/\bar{\Lambda}$ production, we also report results with subtracted feed-down contribution{s} from $\Sigma^0$ and charm. This measurement uses a dataset of 800.4~fb$^{-1}$ collected by the Belle experiment at or near a center-of-mass energy of 10.58 GeV.
We observe a significant polarization that rises with the fractional energy carried by the $\Lambda/\bar{\Lambda}$ hyperon.
\end{abstract}

\maketitle

\setcounter{footnote}{0}
The $\Lambda$ hyperon plays a special role in the study of the spin structure of hadrons due to its self-analyzing weak decay. 
The observation of large transverse polarizations of $\Lambda$ hyperons in unpolarized hadronic collisions over four decades ago~\cite{Bunce:1976yb} was contradictory to the understanding at the time that transverse single-spin asymmetries are suppressed~\cite{Kane:1978nd} in perturbative QCD. This tension helped put in motion a program to study transverse-spin phenomena~\cite{Barone:2010zz}, which has been a major focus of the hadron physics community ever since. Even though there has been tremendous progress in understanding transverse spin phenomena, the original hyperon polarization phenomenon~\cite{Heller1997} still eludes a definitive explanation. A real difficulty is that, in hadronic collisions, it is not possible to disentangle initial-state effects, related to dynamics inside the colliding hadrons, and final-state effects, related to the fragmentation of the partons.

The fragmentation function (FF), describing the production of transversely polarized $\Lambda$ hyperons~\footnote{For simplicity the symbol $\Lambda$ will henceforward be used to refer to both $\Lambda$ and $\bar{\Lambda}$; charge conjugate modes are implied unless explicitly stated.} from unpolarized quarks, is denoted by $D_{1T}^{\perp \Lambda/q}(z, p^2_{\perp})$~\cite{Mulders:1995dh,Anselmino:2000vs}. It depends on the fractional energy, $z$, of the fragmenting quark carried by the observed hyperon and the transverse momentum of the hyperon, $p_{\perp}$, relative to the parent quark. Beyond its connection to the phenomenology of $\Lambda$ production, $D_{1T}^{\perp \Lambda/q}$ has recently been a focus of intense theoretical interest~\cite{Boer:1997mf,Anselmino:2000vs,Boer:2010ya,Wei:2014pma,Chen:2016moq} because it is {time-reversal-odd} (T-odd). It is known that the gauge structure of QCD-universality is modified for the Sivers function, which can be seen as the counterpart of $D_{1T}^{\perp \Lambda/q}$ for the parton distribution function~\cite{Brodsky:2002cx,Collins:2002kn,Ji:2002aa,Belitsky:2002sm,Airapetian:2009ae}. The Sivers function describes the transverse-momentum dependence of unpolarized quarks on the transverse polarization of the parent nucleon, and the predicted sign-change of this function in semi-inclusive deep-inelastic scattering (SIDIS) compared to hadronic collisions has been a focus of several experimental programs~\cite{Perdekamp:2015vwa}. The question of modified universality is equally important for FFs~\cite{Boer:2010ya}, and an extraction of $D_{1T}^{\perp \Lambda/q}$ would be the first measurement of a T-odd and chiral-even FF. 
The chiral-evenness of $D_{1T}^{\perp \Lambda/q}$ arises from the fact that the fragmenting quark is unpolarized, so this function does not have to be sensitive to the spin of the quark. The chiral-evenness is of importance as a test of universality: since the perturbative QCD interactions conserve chirality, chiral-odd functions appear only in combination with other chiral-odd functions, so that the sign is difficult to determine.
We present here the first observation of the transverse polarization of $\Lambda$ hyperons produced in $e^+e^-$ annihilation, from which $D_{1T}^{\perp \Lambda/q}$ can be extracted.

A dataset of 800.4~fb$^{-1}$ at or near $\sqrt{s}= 10.58$ GeV collected by the Belle experiment~\cite{BelleDetector} at the KEKB~\cite{KEKB} $e^+e^-$ collider is used.
For systematic studies and to correct the data for detector effects, Monte Carlo (MC) simulated events are generated using Pythia6.2~\cite{Sjostrand:2000wi} {for fragmentation and Evtgen~\cite{Lange:2001uf} for particle decays}, then processed with a full simulation of the detector response based on a GEANT3~\cite{Brun:1987ma} model of the Belle detector. This measurement considers the processes $e^+e^-\rightarrow \Lambda (\bar{\Lambda})  X$ as well as associated production $e^+e^-\rightarrow \Lambda (\bar{\Lambda})  h^{\pm}  X$, where $h$ denotes a light hadron ($h = \pi, K$) on the opposite side and provides additional information on the fragmenting quark flavor~\cite{Boer:2010ya}.

\begin{figure}
\begin{center}
\includegraphics[width=0.42\textwidth]{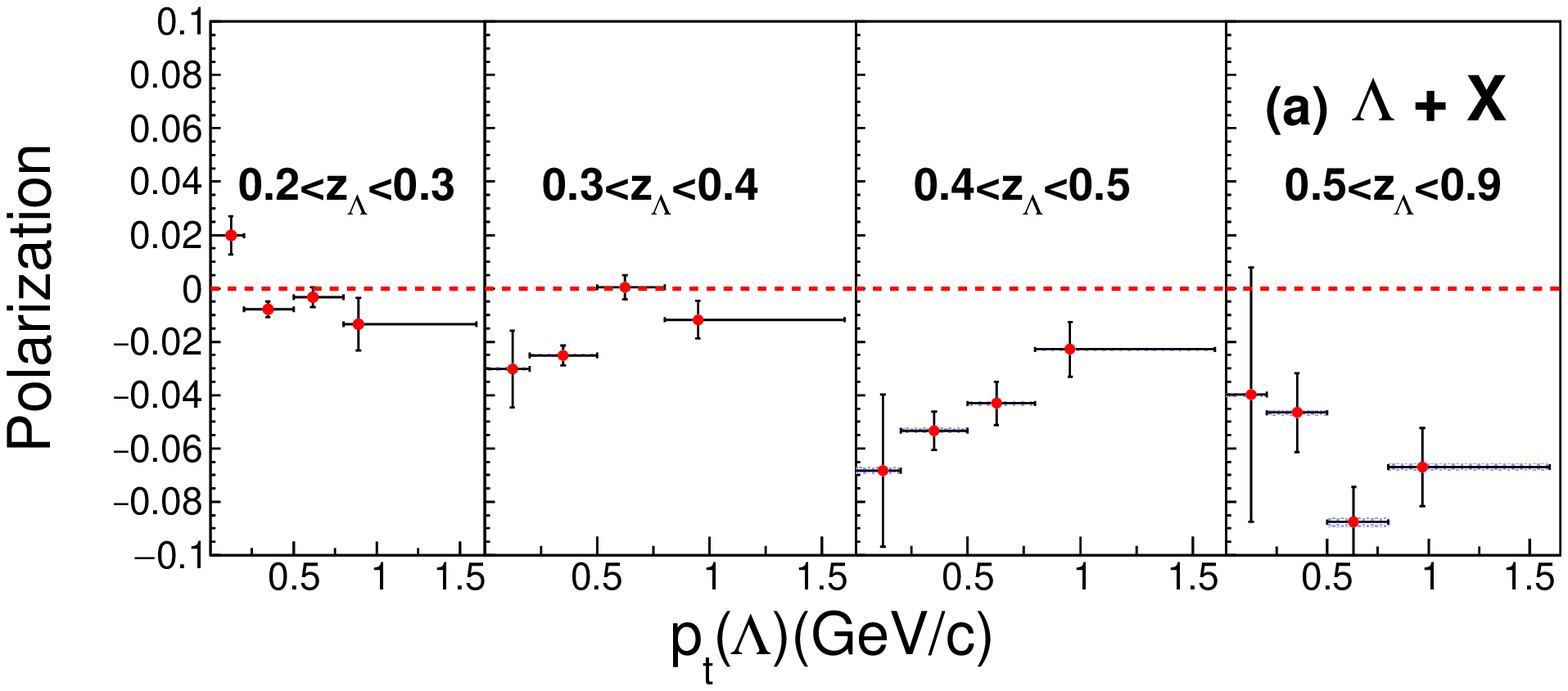}
\includegraphics[width=0.42\textwidth]{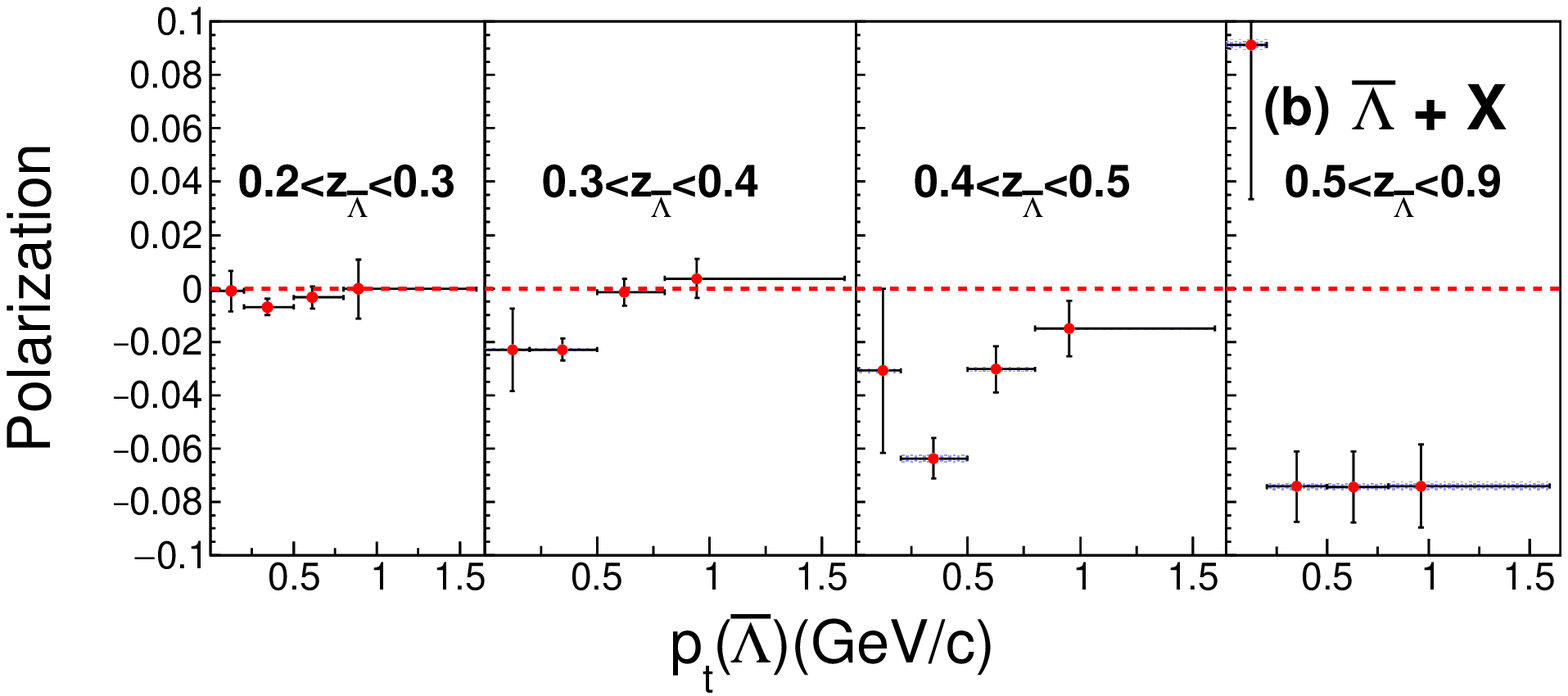}
\caption{Transverse polarization amplitudes of inclusive $\Lambda$'s as a function of $z_\Lambda$ and $p_{\rm t}$ in the thrust frame. 
The top (a) and bottom (b) plots display the results for $\Lambda$ and $\bar{\Lambda}$, respectively. 
The sum of statistical and systematic uncertainties are indicated by the error bars and the shaded areas show the uncertainties from $\alpha$. 
\label{fig:pol_vszpt_sys}}
\end{center}
\end{figure}

Using the event-shape-{variable} thrust, $T$, a sample of light and charm quark fragmentation events, $e^+ e^- \to q\bar{q},\ (q = u,d,s,c)$, is selected~\cite{Seidl:2008xc, Vossen:2011fk}. 
The thrust $T$ is defined in the $e^+e^-$ center-of-mass system as $T= \max \frac{\sum_i |\hat{\mathbf{T}}\cdot {\mathbf{p}}_i|}{\sum_i |{\mathbf{p}}_i|}$.
Here, {$\mathbf{p}_i$} are the momenta of all {detected charged particles and neutral clusters} in the event, 
and {$\hat{\mathbf{T}}$ indicates the unit vector along the thrust axis. 
All charged tracks in the event, with the exception of the $\Lambda$ daughter particles, are required to originate within a region of less than 2.0\,cm in the transverse ($r-\phi$) plane and 4.0\,cm along the beam ($z$) axis with respect to the $e^+e^-$ interaction point (IP). 
We require $T>0.8$, which reduces the contribution of $\Upsilon$ events to less than 1\%. 
In each event, we reconstruct $\Lambda$ candidates from the decay mode $\Lambda \rightarrow p \pi^{-}$. 
The daughter proton and pion are constrained to a decay vertex, and the four-momenta are updated with the vertex constraint.  
The $\Lambda$ candidate is required to have a displaced vertex, consistent with a long-lived particle originating from the IP. 
To further suppress backgrounds, we require the likelihood ($\mathcal{L}$) for one of the daughter particle{s} to be a proton ($p$) by requiring $\mathcal{L}(p)$/($\mathcal{L}(p)$ + $\mathcal{L}(\pi)) >$ 0.6. 
The light hadrons in the associated production are selected in the hemisphere opposite to the $\Lambda$, and are identified using the likelihood ratios $\mathcal{L}(K)$/($\mathcal{L}(K)$ + $\mathcal{L}(p))$ and $\mathcal{L}(K)$/($\mathcal{L}(K)$ + $\mathcal{L}(\pi)$).
The identified muons and electrons are vetoed. In particular, the ratios $\mathcal{L}(K)$/($\mathcal{L}(K)$ + $\mathcal{L}(p)) >$ 0.2 and $\mathcal{L}(K)$/($\mathcal{L}(K)$ + $\mathcal{L}(\pi)) > 0.6$ are required to identify $K^\pm$. And $\mathcal{L}(K)$/($\mathcal{L}(K)$ + $\mathcal{L}(\pi)) < 0.4$ is required to identify $\pi^\pm$.
{Hemispheres are assigned according to the thrust axis, where the axis direction is chosen in such a way that it points into the same hemisphere as the $\Lambda$, that is $\hat{\mathbf{T}} \cdot {\mathbf{p}}_{\Lambda} > 0 $ and $\hat{\mathbf{T}} \cdot {\mathbf{p}}_{h} < 0 $.} The polar angle of the light hadrons ranges from about 0.4~rad to 2.8~rad in the $e^+e^-$ center-of-mass system. 

The transverse momentum of the $\Lambda$, $p_{\rm t}$, is measured with respect to either the thrust axis of the event, or the momentum of the observed hadron in associated production.
We refer to these as the ``thrust frame" and the ``hadron frame", respectively. 
We define the direction $\hat{\mathbf{n}}$ along which the polarization of $\Lambda$ is investigated as $\hat{\mathbf{n}} \propto \hat{\mathbf{m}} \times \hat{\mathbf{p}}_{\Lambda}$, where $\hat{\mathbf{m}}$ is equal to $\hat{\mathbf{T}}$ ($\mathcal{-}\hat{\mathbf{p}}_h$) in the thrust (hadron) frame.
Given a transverse polarization $P$ of the $\Lambda$, the distribution of protons from the $\Lambda$ decays is given by
\begin{equation}
\label{eq:main}
\frac{1}{N} \frac{dN}{d \rm{cos}\theta} = 1 + \alpha P {\rm cos}\theta,
\end{equation}
where $N$ is the total signal yield, $\theta$ is the angle between $\hat{\mathbf{n}}$ and the proton momentum in the $\Lambda$ rest frame, 
and $\alpha = 0.642 ~ \pm ~ 0.013 $ is the world average value of the parity-violating decay asymmetry for the $\Lambda$~\cite{PDG}. 
Assuming $CP$ conservation, the value of $\alpha$ for the $\bar{\Lambda}$ decay is of the same magnitude as for the $\Lambda$ with an opposite sign.

The $\Lambda$ signal is clearly observed in the invariant mass ($M_{p\pi^-}$) spectrum, and the purity of the $\Lambda$ {($\bar{\Lambda}$}) is about 91\% (93\%). {A linear average of the $\cost$ distributions of events in the sideband regions, [1.103, 1.108]\,GeV/$c^2$ and [1.123, 1.128]\,GeV/$c^2$, is subtracted from that in the signal region, [1.11, 1.12]\,GeV/$c^2$, to exclude background contributions.} The transverse polarization of the $\Lambda$ is investigated as a function of $z_{\Lambda}$ and $p_{\rm t}$, where $z_{\Lambda} = 2E_{\Lambda}/\sqrt{s}$.
Four $z_\Lambda$ bins with boundaries at $z_{\Lambda} = [0.2, 0.3, 0.4, 0.5, 0.9]$, four $p_{\rm t}$ bins with boundaries at $p_{\rm t} = [0.0, 0.2, 0.5, 0.8, 1.6]\,{\rm{GeV}}/c$, {and five $\cost$ bins} are adopted in the thrust frame.
{To correct for detector inefficiencies, the dependence of the efficiency on $\cos\theta$ is derived from MC.}
Also, due to the smearing in the reconstruction of $z_{\Lambda}$, $p_{\rm t}$ and $\cost$, bin-to-bin migrations are expected. 
Based on MC, we find {that} in the thrust frame, the bin migration is dominated by the smearing in $p_{\rm t}$, which is caused by the resolution of the thrust axis. 
Depending on {the} $z_{\Lambda}$ range, {between 2\% and 35\% of} the events are falsely reconstructed in the adjacent $p_{\rm t}$ bins.
An unfolding procedure based on the singular value decomposition (svd) is used to correct the $z_{\Lambda}$, $p_{\rm t}$, and $\cost$ smearing {and detector efficiencies} simultaneously~\cite{svdunfold}.
The sideband subtracted $\cos\theta$ distributions are used as input in the svd unfolding. The response matrix is estimated from MC. The unfolded $\cost$ distributions are then self-normalized: $R(\theta) = N(\theta)/{\langle N \rangle}$, where $\langle N \rangle$ denotes the averaged number of events in each $\cost$ bin. The normalized $\cost$ distributions are then individually fit using the function $1 + f_{0} {\rm cos}\theta$, where $f_{0}$ is a free parameter. 
The magnitude of the polarization is $P= f_{0}/\alpha$. 
The obtained {polarizations} are displayed in Fig.~\ref{fig:pol_vszpt_sys}.

A significant transverse polarization is observed. In general, the magnitude of the polarization rises with $z_\Lambda$. 
The $p_{\rm t}$ behavior is more complex and depends on the $z_\Lambda$ range. 
For $z_\Lambda>$ 0.5, where the $\Lambda$ is the leading particle, and for {$z_\Lambda <$ 0.3}, we observe rising asymmetries with $p_{\rm t}$. 
In contrast, for intermediate $z_\Lambda$, the dependence {seems to be} reversed. 
{This behavior may be caused by different quark-flavor contributions in the different [$z_\Lambda$, $p_{\rm t}$] regions, as different quark flavors can {give} rise to different polarizations and kinematic dependencies.}}
{Based on MC~\cite{SM} studies, for $\Lambda$~\cite{state},
in the {highest} $z_\Lambda$ bin, {the} $s$ quark contribution is dominant. In the intermediate two $z_\Lambda$ bins, there is less $s$ quark contribution compared to the {highest} $z_\Lambda$ bin.
The contribution of $u$ quark{s}, which could produce polarization with a different sign compared to $s$ quarks, might cancel the $s$ quark contribution and cause the reversed $p_t$ dependence.
{However}, it should also be noted that there is a larger charm contribution in the two intermediate $z_\Lambda$ bins~\cite{SM}}.

Considering associated production of a light hadron on the opposite side, four $z_{h}$ bins with boundaries at $z_{h} = [0.2, 0.3, 0.4, 0.5, 0.9]$ are adopted, where $z_{h} = 2E_{h}/\sqrt{s}$. In the hadron frame, the detector smearing effects are found {to be} negligible because of the much better resolution of $\mathcal{-}\hat{\mathbf{p}}_h$ compared to that of $\hat{\mathbf{T}}$. Also, less than 5\% of events are falsely reconstructed in the wrong $z_\Lambda$ or $z_h$ bins. Thus, svd unfolding is not applied here. The {{efficiency}}-corrected $\cos\theta$ distributions are fit {in the same way} as those in the thrust frame.
{Due to particle mis-identifications, the purity of the $\pi^+$ ($\pi^-$) is about 91.8\% (94.8\%) and that of $K^+$ ($K^-$) is 87.4\% (69.8\%), based on MC. 
{The contributions from mis-identified $h^{\pm}$ are included in the results without further correction.} The amplitudes of the transverse polarization of $\Lambda$ hyperons as a function of $z_{\Lambda}$ and $z_{h}$ calculated in the hadron frame are shown in Fig.~\ref{fig:pol_zlzh_sys_href}. 
These results can give additional insight into the quark flavor fragmenting into the $\Lambda$.
In particular, in the low $z_\Lambda$ region, the polarization in $\Lambda  h^+ X$ and $\Lambda  h^- X $ is significantly different, 
even showing opposite sign and a magnitude that increases with higher $z_h$. 
In contrast, in the region $z_\Lambda > 0.5$, the differences between $\Lambda  h^+ X$ and $\Lambda  h^- X$ are modest, although deviations can still be seen.

We investigate the flavor of the (anti-)quark going into the same hemisphere with the $\Lambda$ particles using MC. 
We find that the flavor tag of the light hadron depends on $z_h$ and $z_\Lambda${~\cite{SM}}. 
At low $z_\Lambda$~\cite{state}, the contributions of the various quark flavors for $\Lambda$ are nearly charge symmetric in processes $\Lambda  h^+  X $ and $\Lambda  h^-  X $.
In general, the results suggest that the $\Lambda$ polarization from $s$ quark fragmentation is negative because, in $\Lambda  K^+  X$ at high $z_\Lambda$, where ${s}$ to $\Lambda$ fragmentation absolutely dominates, the observed asymmetries are negative. In $\Lambda  \pi^-  X$ and $\Lambda  K^-  X$ at low $z_\Lambda$, ${u}$ to $\Lambda$ fragmentation dominates, and the observed positive asymmetries suggest that the $u$ quark fragmentation to $\Lambda$ is positive. In $\Lambda  \pi^-  X$ and $\Lambda  K^-  X$ at high $z_\Lambda$, there is a larger contribution from $s$ compared to low $z_\Lambda$, resulting in negative polarizations. 
For $\Lambda  \pi^+  X$ at low $z_\Lambda$, {$\bar{u}$ fragmenting into a $\Lambda$} dominates,  
and the observed polarizations are negative. At high $z_\Lambda$, ${s}$ fragmenting into $\Lambda$ is dominant, resulting in negative polarization.
The sign of the $\Lambda$ polarization fragmenting from $d$ quark{s} is not well determined.
\begin{figure}
\begin{center}
\includegraphics[width=0.42\textwidth]{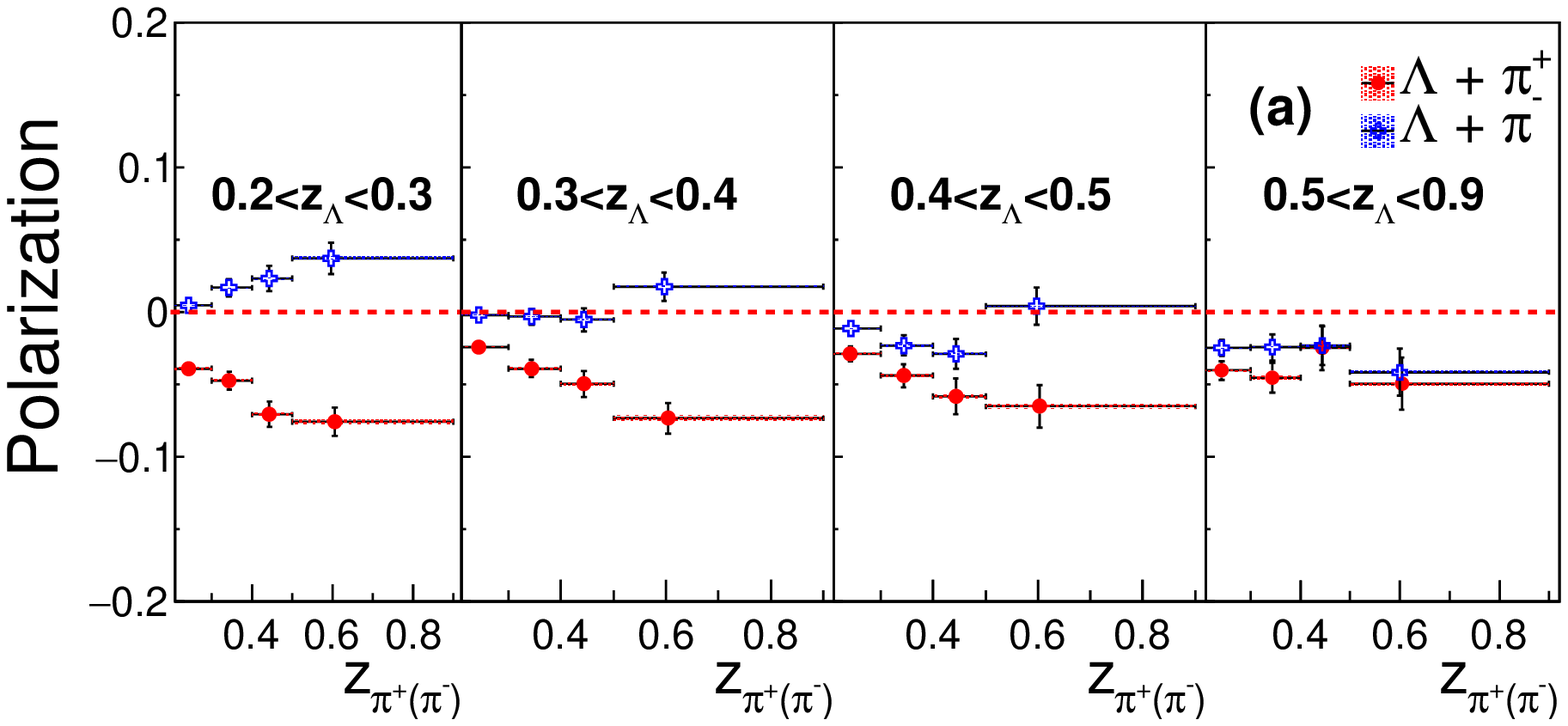}
\includegraphics[width=0.42\textwidth]{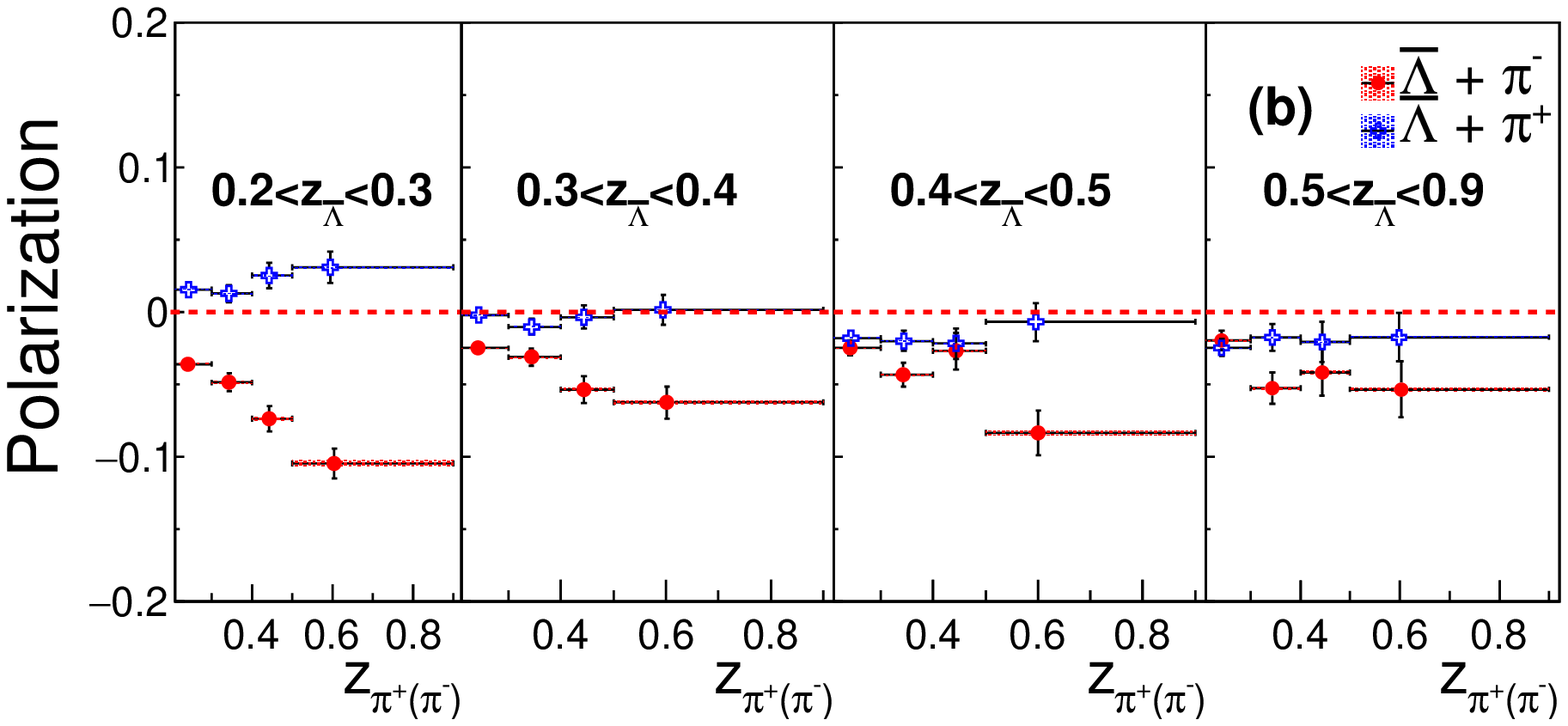}
\includegraphics[width=0.42\textwidth]{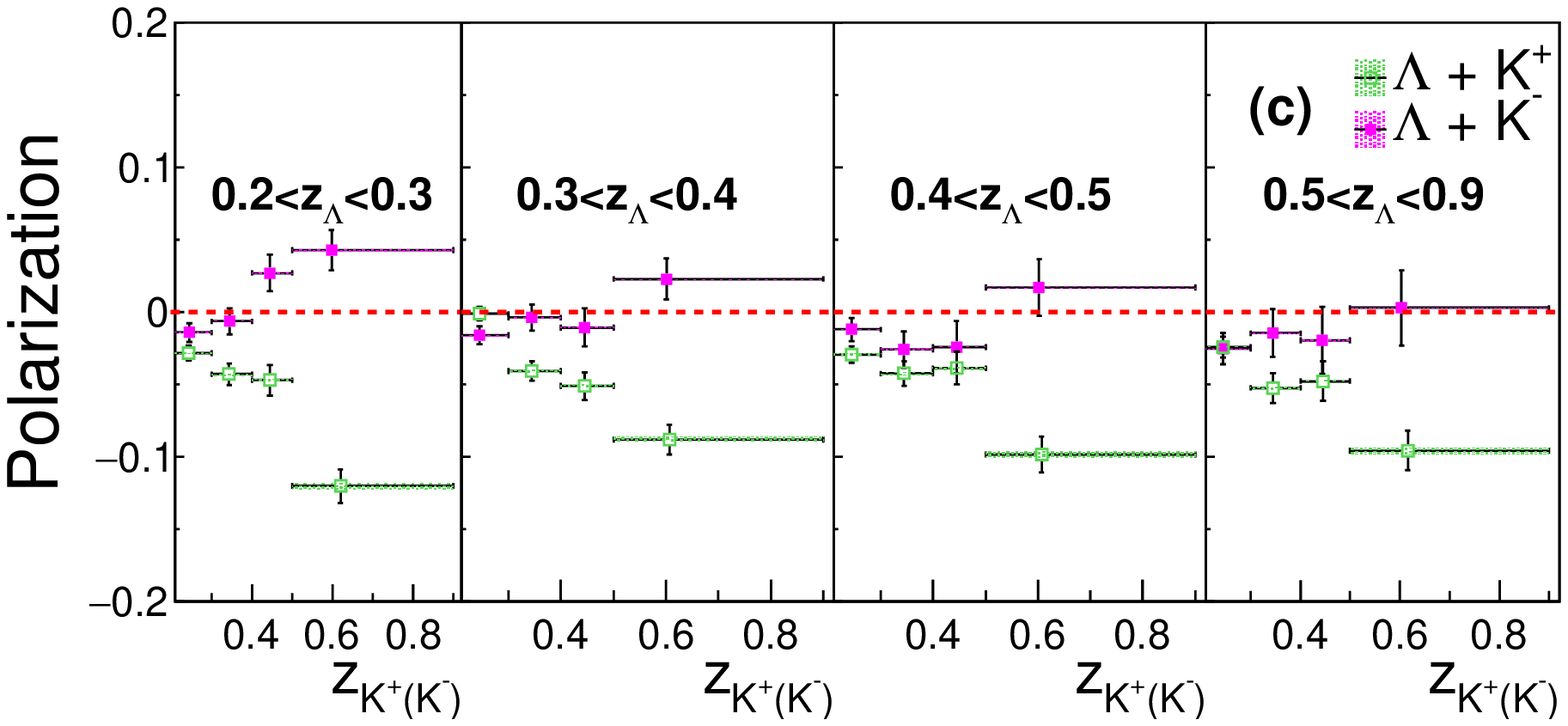}
\includegraphics[width=0.42\textwidth]{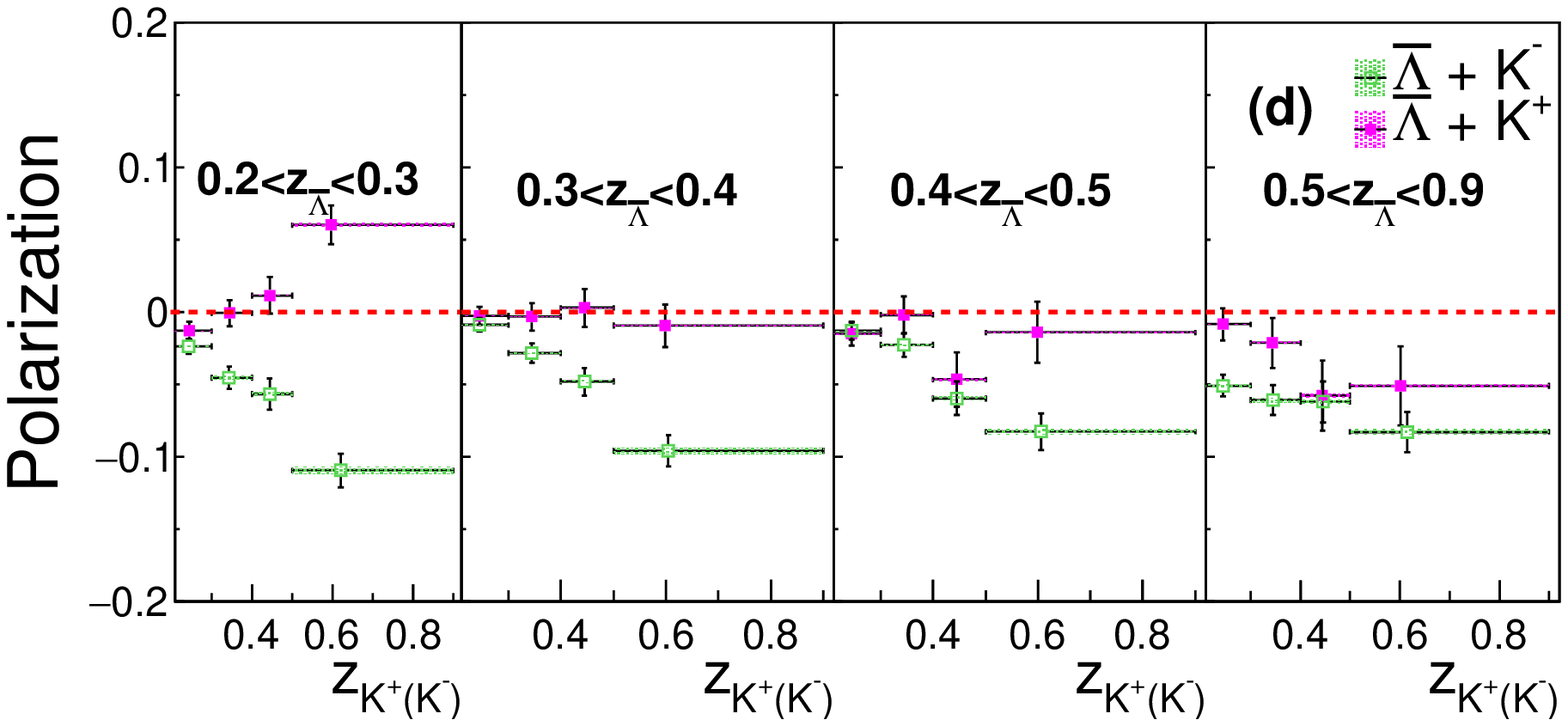}
\caption{Transverse polarizations of $\Lambda$'s observed in 
$\Lambda  \pi^{\pm}  X$ (a), $\Lambda  K^{\pm}  X$ (b), $\bar{\Lambda}  \pi^{\pm}  X$ (c) and $\bar{\Lambda}  K^{\pm}  X$ (d), as a function of $z_\Lambda$ and $z_{h}$ in the hadron frame. 
The different panels show the different $z_\Lambda$ regions as labeled on the plots. Error bars indicate the sum of statistical and systematic uncertainties added in quadrature. The shaded areas show the uncertainties from $\alpha$.}
\label{fig:pol_zlzh_sys_href}
\end{center}
\end{figure}

The results presented in Fig.~\ref{fig:pol_vszpt_sys} and Fig.~\ref{fig:pol_zlzh_sys_href} show the transverse polarization for inclusive $\Lambda$ particles, including those directly-produced from $q\bar{q}$ fragmentations and those indirectly-produced from decays. 
{Based on MC, about 30\% of $\Lambda$ candidates come from charm, mainly via $c\rightarrow\Lambda_c$, and in light quarks ($uds$) about 20\% of the $\Lambda$ candidates {come} from $\Sigma^0$ and 10\% from $\Xi$ decays.
We note that the strong decays, such as that of {$\Sigma^*$}, are considered {as} part of the fragmentation function.
{The charm is expected to be different from light quarks because it is much heavier, thus we {need to also separately correct for the charm contribution}. {To study direct fragmentation of light quarks into $\Lambda$ hyperons, also the contributions from $\Sigma^0$ and $\Xi$ decays need to be taken into account.} 
We analyzed $\Sigma^0$-, $\Xi$- and $D$-enhanced samples.
The $D$-enhanced sample serves as a tag for charm events. 
The $\Sigma^0$ is reconstructed from $\Sigma^0 \rightarrow \Lambda  \gamma$, which practically saturates the branching fraction of the $\Sigma^0$~\cite{PDG}, and the $\Xi$ is reconstructed from $\Xi^{-} \rightarrow \Lambda  \pi^-$, which also saturates the branching fraction of the $\Xi$, while $D$ mesons are reconstructed using $D^0 \rightarrow K^-\pi^+ $ and $D^+ \rightarrow K^-\pi^+\pi^+$ modes. 
No hemisphere requirement is imposed on the $\Sigma^0$ or $\Xi$ candidates. $D$ candidates are required to be in the opposite hemisphere.
{{An} invariant-mass window is required to select the $\Sigma^0$($\Xi$, $D$)-enhanced sample. 
Events without $\Sigma^0$($D$) candidates are referred {to} as the $\Sigma^0$($D$)-suppressed samples.} 
The $\Xi$-enhanced sample is found having consistent polarizations with the nominal sample within statistical uncertainties. Also, given the relatively smaller contribution, $\Xi$ is considered {as} part of the signal.
We correct {for} the feed-down from charm and $\Sigma^0$ in light quarks. The measured polarization can be expressed as:
\begin{equation}~\label{eq:bkgunfold}
P^{\rm mea} = (1 - \sum_{i}F_{i}) P^{{\rm prompt}} + \sum_{i}F_{i}P_{i},
\end{equation}
where $P^{\rm prompt}$ is the polarization of signal $\Lambda$ particles from light quarks, $P_{i}$ is the polarization associated with the $i^{\rm th}$ feed-down process and $F_i$ is the fraction of the $i^{\rm th}$ process.
The $F_{i}$ are estimated from MC but scaled according to measured cross sections for $\Sigma^0$~\cite{Niiyama:2017wpp} and $D$ mesons. {We have three main processes for feed-down production of $\Lambda$ particles: from $\Sigma^0$ decays in $uds$, from $\Sigma^0$ decays in charm, and from other sources in charm.} {We have four measurements of polarizations with different $F_{i}$ using four samples: $\Sigma^0$-enhanced-$D$-enhanced, $\Sigma^0$-enhanced-$D$-suppressed, $\Sigma^0$-suppressed-$D$-enhanced and $\Sigma^0$-suppressed-$D$-suppressed.} 
Then the feed-down-corrected polarizations are determined by solving {Eq.~(\ref{eq:bkgunfold})} for the five $z_{\Lambda}$ bins in the thrust frame. We cannot consider the transverse momentum dependence {in either reference frame} due to limited statistics. {The bin-to-bin migrations are not significant between different $z_\Lambda$ bins, and hence the svd unfolding is not applied here. A factor estimated from MC, which ranges from 1.1 to 1.3, is used to correct for the detector smearing effects on the cos$\theta$ distributions.} 
The feed-down-corrected results are shown in Fig.~\ref{fig:bkgunfolded}.
Given the large uncertainties, no strong conclusion can be drawn from the results for $\Lambda$ from charm production or $\Sigma^0$ decays.

\begin{figure}
\begin{center}
\includegraphics[width=0.42\textwidth]{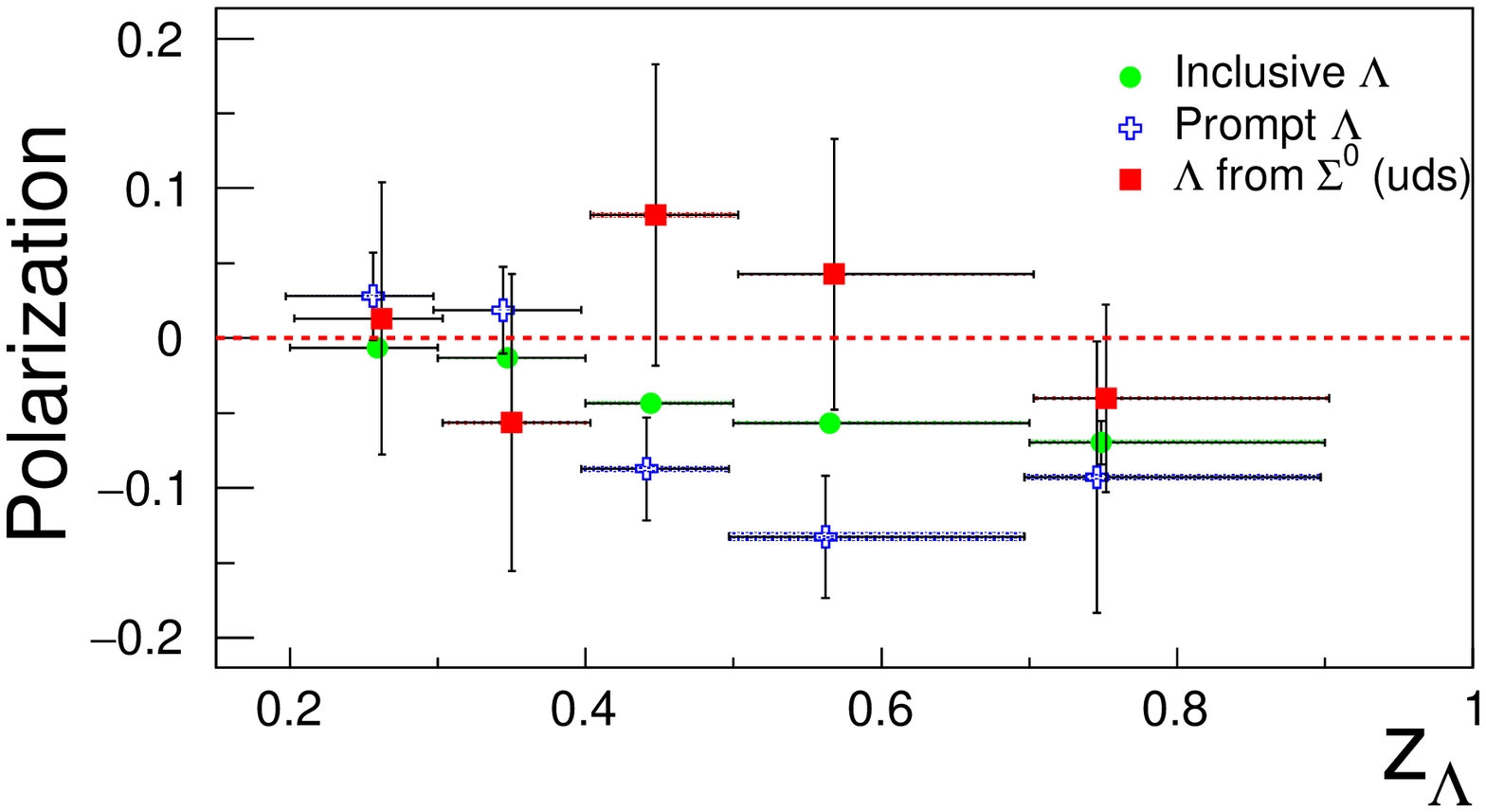}
\includegraphics[width=0.42\textwidth]{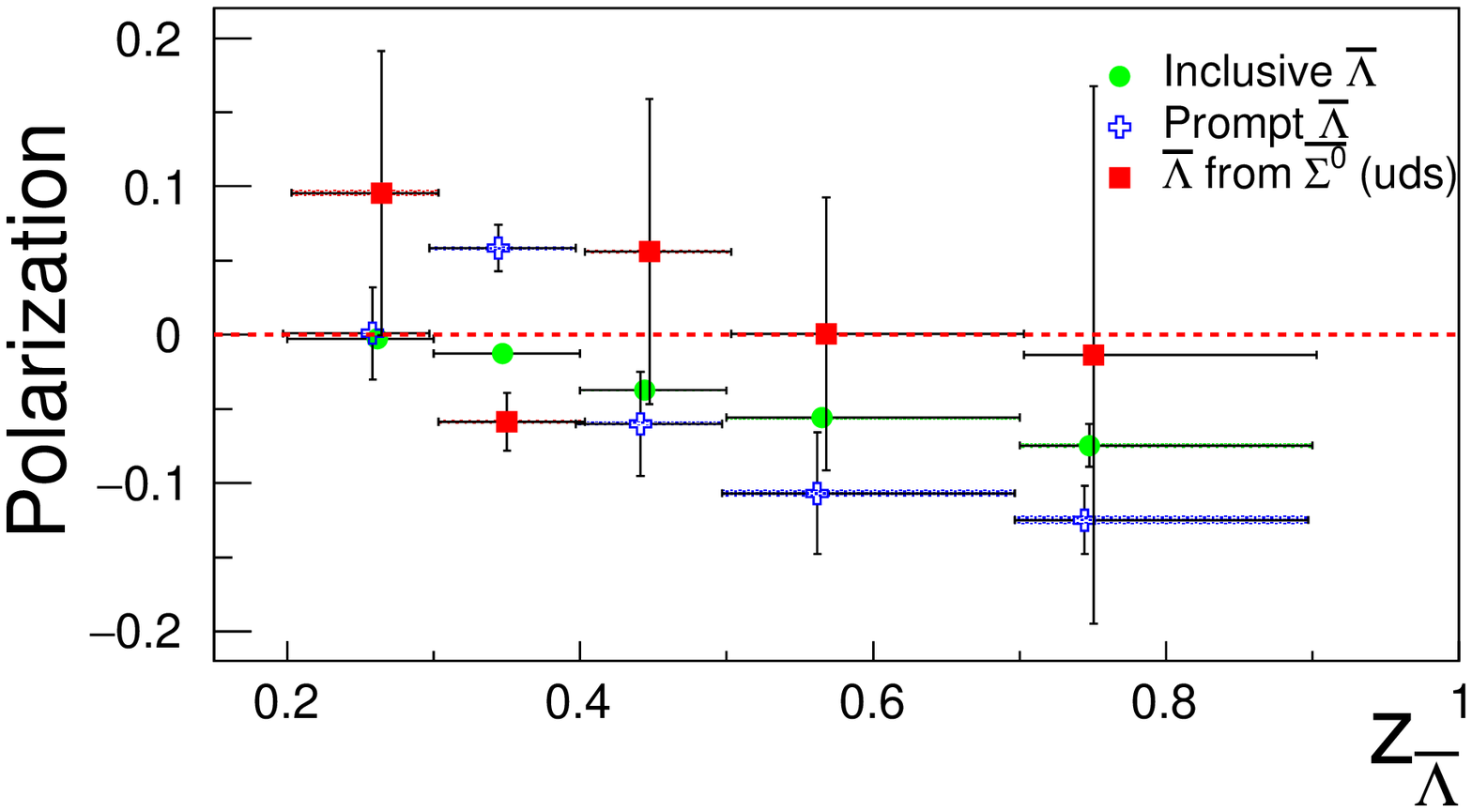}
\caption{The unfolded transverse polarizations of prompt $\Lambda$'s {from $uds$ fragmentation} (blue crosses) and $\Lambda$'s from $\Sigma^0 \rightarrow \Lambda \gamma$ decays (red squares) {in $uds$ fragmentation}, compared to the original polarizations observed for inclusive $\Lambda$'s (green dots), as a function of $z_\Lambda$ in the thrust frame. Error bars show the statistical and systematic uncertainties added in quadrature. The shaded areas show the uncertainties from $\alpha$.} 
\label{fig:bkgunfolded}
\end{center}
\end{figure}

Systematic uncertainties from the sideband subtraction are estimated by varying the scale factor of events in {the} sidebands. When the shape used to describe the background contributions under the $M_{p\pi^-}$ mass peak is changed from a first-order to a second-order polynomial function, the obtained scale factor increases from 1.0 to 1.3.
The resulting variations on the polarizations range from 0.000 to 0.002 for different bins {and} are assigned as systematic uncertainties.

The covariance matrix due to the MC statistics in the svd unfolding is assigned as a systematic uncertainty. The resultant uncertainties range from 0.001 to 0.016 for different bins.
The reconstructed $z_\Lambda$ and $p_{\rm t}$ distributions are found to be slightly different in data and MC.
The response matrix of the MC thus is varied according to these differences and the changes on the obtained polarizations, which range from 0.000 to 0.033 for different bins, are assigned as systematic uncertainties. For the feed-down-corrected results shown in Fig.~\ref{fig:bkgunfolded}, the uncertainties of the correction factor for detector smearing from limited MC statistics are assigned as systematic uncertainties.

We estimate the systematics from possible non-linear $\cost$ contributions by adding a second-order term to the fitting model,
described as {$f_{0} + f_{1}\cost + f_{2}{\rm{cos}}^{2}\theta$}, where $f_{0}$, $f_{1}$ and $f_{2}$ are free parameters. 
The differences in the extracted polarizations ($f_{1}/\alpha$) from the nominal values, ranging from 0.0000 to 0.0003, are assigned as systematic uncertainties.  
All systematic uncertainties are then added in quadrature. 
In addition, the scale uncertainty from the decay parameter $\alpha$~\cite{PDG} is assigned and displayed separately as shaded areas in the figures.

We perform two checks to verify that our measurement is not biased.
First, the reference axis is replaced by $\hat{\mathbf{n}}' \equiv \pm \hat{\mathbf{p}}_{\Lambda} \times \hat{\mathbf{n}}$, which is still normal to {the} $\Lambda$ direction but in the $\Lambda$ production plane. Second, we use event mixing by reconstructing $\Lambda$ candidates using a proton and a pion from different events. No significant bias is observed.

In summary, we have studied the transverse polarization of $\Lambda (\bar{\Lambda})$ in the inclusive processes $e^+e^-\to \Lambda (\bar{\Lambda})  X$ 
and $e^+e^-\to \Lambda (\bar{\Lambda})   h^{\pm}  X$ with the data collected by Belle. 
A significant transverse polarization is observed, which is the first such observation in $e^+e^-$ annihilation. Its magnitude as a function of $z_\Lambda$ and $p_{\rm t}$ is presented, and increases with $z_\Lambda$ as predicted~\cite{Boer:2010ya}. 
The results are consistent between inclusive $\Lambda$ and $\bar{\Lambda}$ production.
By selecting an identified light hadron in the opposite hemisphere, we obtain sensitivity to the flavor dependence of the observed polarization.  
{Strong flavor dependences are} seen in the $\Lambda (\bar{\Lambda})  h^{\pm} X $ measurements.
Our results} suggest positive polarization for $u$ ($\bar{u}$) quark fragmentation to a $\Lambda$ ($\bar{\Lambda}$) and negative polarization for $s$ ($\bar{s}$) quark fragmentation to a $\Lambda$ ($\bar{\Lambda}$). 
{{A} conclusive understanding} needs more dedicated studies with theoretical calculations.
Furthermore, we attempt to separate the contributions for directly-produced $\Lambda$ particles from light quarks and those from charm and $\Sigma^0$ decays.
The results presented in this Letter provide rich information about the transverse polarization of $\Lambda$ hyperons and will further contribute to the understanding of the fragmentation processes in $\Lambda$ production. {{These results will also be useful to test the universality of T-odd FFs, in combination with data from hadron collisions~\cite{Bunce:1976yb,Lundberg:1989hw, Ramberg:1994tk,Abt:2006da,Fanti:1998px,ATLAS:2014ona} and future SIDIS data.
\begin{acknowledgments} 
We thank the KEKB group for excellent operation of the
accelerator; the KEK cryogenics group for efficient solenoid
operations; and the KEK computer group, the NII, and 
PNNL/EMSL for valuable computing and SINET5 network support.  
We acknowledge support from MEXT, JSPS and Nagoya's TLPRC (Japan);
ARC (Australia); FWF (Austria); NSFC and CCEPP (China); 
MSMT (Czechia); CZF, DFG, EXC153, and VS (Germany);
DST (India); INFN (Italy); 
MOE, MSIP, NRF, RSRI, FLRFAS project and GSDC of KISTI and KREONET/GLORIAD (Korea);
MNiSW and NCN (Poland); MES (Russia); ARRS (Slovenia);
IKERBASQUE and MINECO (Spain); 
SNSF (Switzerland); MOE and MOST (Taiwan); and DOE and NSF (USA).
\end{acknowledgments}
\bibliography{lambda}

\begin{thebibliography}{34}%
\makeatletter
\providecommand \@ifxundefined [1]{%
 \@ifx{#1\undefined}
}%
\providecommand \@ifnum [1]{%
 \ifnum #1\expandafter \@firstoftwo
 \else \expandafter \@secondoftwo
 \fi
}%
\providecommand \@ifx [1]{%
 \ifx #1\expandafter \@firstoftwo
 \else \expandafter \@secondoftwo
 \fi
}%
\providecommand \natexlab [1]{#1}%
\providecommand \enquote  [1]{``#1''}%
\providecommand \bibnamefont  [1]{#1}%
\providecommand \bibfnamefont [1]{#1}%
\providecommand \citenamefont [1]{#1}%
\providecommand \href@noop [0]{\@secondoftwo}%
\providecommand \href [0]{\begingroup \@sanitize@url \@href}%
\providecommand \@href[1]{\@@startlink{#1}\@@href}%
\providecommand \@@href[1]{\endgroup#1\@@endlink}%
\providecommand \@sanitize@url [0]{\catcode `\\12\catcode `\$12\catcode
  `\&12\catcode `\#12\catcode `\^12\catcode `\_12\catcode `\%12\relax}%
\providecommand \@@startlink[1]{}%
\providecommand \@@endlink[0]{}%
\providecommand \url  [0]{\begingroup\@sanitize@url \@url }%
\providecommand \@url [1]{\endgroup\@href {#1}{\urlprefix }}%
\providecommand \urlprefix  [0]{URL }%
\providecommand \Eprint [0]{\href }%
\providecommand \doibase [0]{http://dx.doi.org/}%
\providecommand \selectlanguage [0]{\@gobble}%
\providecommand \bibinfo  [0]{\@secondoftwo}%
\providecommand \bibfield  [0]{\@secondoftwo}%
\providecommand \translation [1]{[#1]}%
\providecommand \BibitemOpen [0]{}%
\providecommand \bibitemStop [0]{}%
\providecommand \bibitemNoStop [0]{.\EOS\space}%
\providecommand \EOS [0]{\spacefactor3000\relax}%
\providecommand \BibitemShut  [1]{\csname bibitem#1\endcsname}%
\let\auto@bib@innerbib\@empty
\bibitem [{\citenamefont {Bunce}\ \emph {et~al.}(1976)\citenamefont {Bunce}
  \emph {et~al.}}]{Bunce:1976yb}%
  \BibitemOpen
  \bibfield  {author} {\bibinfo {author} {\bibfnamefont {G.}~\bibnamefont
  {Bunce}} \emph {et~al.},\ }\href {\doibase 10.1103/PhysRevLett.36.1113}
  {\bibfield  {journal} {\bibinfo  {journal} {Phys. Rev. Lett.}\ }\textbf
  {\bibinfo {volume} {36}},\ \bibinfo {pages} {1113} (\bibinfo {year}
  {1976})}\BibitemShut {NoStop}%
\bibitem [{\citenamefont {Kane}\ \emph {et~al.}(1978)\citenamefont {Kane},
  \citenamefont {Pumplin},\ and\ \citenamefont {Repko}}]{Kane:1978nd}%
  \BibitemOpen
  \bibfield  {author} {\bibinfo {author} {\bibfnamefont {G.~L.}\ \bibnamefont
  {Kane}}, \bibinfo {author} {\bibfnamefont {J.}~\bibnamefont {Pumplin}}, \
  and\ \bibinfo {author} {\bibfnamefont {W.}~\bibnamefont {Repko}},\ }\href
  {\doibase 10.1103/PhysRevLett.41.1689} {\bibfield  {journal} {\bibinfo
  {journal} {Phys. Rev. Lett.}\ }\textbf {\bibinfo {volume} {41}},\ \bibinfo
  {pages} {1689} (\bibinfo {year} {1978})}\BibitemShut {NoStop}%
\bibitem [{\citenamefont {Barone}\ \emph {et~al.}(2010)\citenamefont {Barone},
  \citenamefont {Bradamante},\ and\ \citenamefont {Martin}}]{Barone:2010zz}%
  \BibitemOpen
  \bibfield  {author} {\bibinfo {author} {\bibfnamefont {V.}~\bibnamefont
  {Barone}}, \bibinfo {author} {\bibfnamefont {F.}~\bibnamefont {Bradamante}},
  \ and\ \bibinfo {author} {\bibfnamefont {A.}~\bibnamefont {Martin}},\ }\href
  {\doibase 10.1016/j.ppnp.2010.07.003} {\bibfield  {journal} {\bibinfo
  {journal} {Prog. Part. Nucl. Phys.}\ }\textbf {\bibinfo {volume} {65}},\
  \bibinfo {pages} {267} (\bibinfo {year} {2010})}\BibitemShut {NoStop}%
\bibitem [{Hel()}]{Heller1997}%
  \BibitemOpen
  \href@noop {} {\ }\bibinfo {note} {For a review of the data, see, e.g., K.
  Heller, in Proceedings of the 12th International Symposium on Spin Physics,
  Amsterdam, 1996 (World Scientific, Singapore, 1997).}\BibitemShut {Stop}%
\bibitem [{Note1()}]{Note1}%
  \BibitemOpen
  \bibinfo {note} {For simplicity the symbol $\Lambda $ will henceforward be
  used to refer to both $\Lambda $ and $\protect \mathaccentV {bar}016{\Lambda
  }$; charge conjugate modes are implied unless explicitly stated.}\BibitemShut
  {Stop}%
\bibitem [{\citenamefont {Mulders}\ and\ \citenamefont
  {Tangerman}(1996)}]{Mulders:1995dh}%
  \BibitemOpen
  \bibfield  {author} {\bibinfo {author} {\bibfnamefont {P.~J.}\ \bibnamefont
  {Mulders}}\ and\ \bibinfo {author} {\bibfnamefont {R.~D.}\ \bibnamefont
  {Tangerman}},\ }\href@noop {} {\bibfield  {journal} {\bibinfo  {journal}
  {Nucl. Phys.}\ }\textbf {\bibinfo {volume} {B461}},\ \bibinfo {pages} {197}
  (\bibinfo {year} {1996})},\ \bibinfo {note} {[Erratum: Nucl. Phys.
  $\bf{B484}$, 538 (1997)]}\BibitemShut {NoStop}%
\bibitem [{\citenamefont {Anselmino}\ \emph {et~al.}(2001)\citenamefont
  {Anselmino}, \citenamefont {Boer}, \citenamefont {D'Alesio},\ and\
  \citenamefont {Murgia}}]{Anselmino:2000vs}%
  \BibitemOpen
  \bibfield  {author} {\bibinfo {author} {\bibfnamefont {M.}~\bibnamefont
  {Anselmino}}, \bibinfo {author} {\bibfnamefont {D.}~\bibnamefont {Boer}},
  \bibinfo {author} {\bibfnamefont {U.}~\bibnamefont {D'Alesio}}, \ and\
  \bibinfo {author} {\bibfnamefont {F.}~\bibnamefont {Murgia}},\ }\href
  {\doibase 10.1103/PhysRevD.63.054029} {\bibfield  {journal} {\bibinfo
  {journal} {Phys. Rev. D}\ }\textbf {\bibinfo {volume} {63}},\ \bibinfo
  {pages} {054029} (\bibinfo {year} {2001})}\BibitemShut {NoStop}%
\bibitem [{\citenamefont {Boer}\ \emph {et~al.}(1997)\citenamefont {Boer},
  \citenamefont {Jakob},\ and\ \citenamefont {Mulders}}]{Boer:1997mf}%
  \BibitemOpen
  \bibfield  {author} {\bibinfo {author} {\bibfnamefont {D.}~\bibnamefont
  {Boer}}, \bibinfo {author} {\bibfnamefont {R.}~\bibnamefont {Jakob}}, \ and\
  \bibinfo {author} {\bibfnamefont {P.~J.}\ \bibnamefont {Mulders}},\ }\href
  {\doibase 10.1016/S0550-3213(97)00456-2} {\bibfield  {journal} {\bibinfo
  {journal} {Nucl. Phys.}\ }\textbf {\bibinfo {volume} {B504}},\ \bibinfo
  {pages} {345} (\bibinfo {year} {1997})}\BibitemShut {NoStop}%
\bibitem [{\citenamefont {Boer}\ \emph {et~al.}(2010)\citenamefont {Boer},
  \citenamefont {Kang}, \citenamefont {Vogelsang},\ and\ \citenamefont
  {Yuan}}]{Boer:2010ya}%
  \BibitemOpen
  \bibfield  {author} {\bibinfo {author} {\bibfnamefont {D.}~\bibnamefont
  {Boer}}, \bibinfo {author} {\bibfnamefont {Z.-B.}\ \bibnamefont {Kang}},
  \bibinfo {author} {\bibfnamefont {W.}~\bibnamefont {Vogelsang}}, \ and\
  \bibinfo {author} {\bibfnamefont {F.}~\bibnamefont {Yuan}},\ }\href {\doibase
  10.1103/PhysRevLett.105.202001} {\bibfield  {journal} {\bibinfo  {journal}
  {Phys. Rev. Lett.}\ }\textbf {\bibinfo {volume} {105}},\ \bibinfo {pages}
  {202001} (\bibinfo {year} {2010})}\BibitemShut {NoStop}%
\bibitem [{\citenamefont {Wei}\ \emph {et~al.}(2015)\citenamefont {Wei},
  \citenamefont {Chen}, \citenamefont {Song},\ and\ \citenamefont
  {Liang}}]{Wei:2014pma}%
  \BibitemOpen
  \bibfield  {author} {\bibinfo {author} {\bibfnamefont {S.~Y.}\ \bibnamefont
  {Wei}}, \bibinfo {author} {\bibfnamefont {K.~B.}\ \bibnamefont {Chen}},
  \bibinfo {author} {\bibfnamefont {Y.~K.}\ \bibnamefont {Song}}, \ and\
  \bibinfo {author} {\bibfnamefont {Z.~T.}\ \bibnamefont {Liang}},\ }\href
  {\doibase 10.1103/PhysRevD.91.034015} {\bibfield  {journal} {\bibinfo
  {journal} {Phys. Rev. D}\ }\textbf {\bibinfo {volume} {91}},\ \bibinfo
  {pages} {034015} (\bibinfo {year} {2015})}\BibitemShut {NoStop}%
\bibitem [{\citenamefont {Chen}\ \emph {et~al.}(2016)\citenamefont {Chen},
  \citenamefont {Yang}, \citenamefont {Wei},\ and\ \citenamefont
  {Liang}}]{Chen:2016moq}%
  \BibitemOpen
  \bibfield  {author} {\bibinfo {author} {\bibfnamefont {K.~B.}\ \bibnamefont
  {Chen}}, \bibinfo {author} {\bibfnamefont {W.~H.}\ \bibnamefont {Yang}},
  \bibinfo {author} {\bibfnamefont {S.~Y.}\ \bibnamefont {Wei}}, \ and\
  \bibinfo {author} {\bibfnamefont {Z.~T.}\ \bibnamefont {Liang}},\ }\href
  {\doibase 10.1103/PhysRevD.94.034003} {\bibfield  {journal} {\bibinfo
  {journal} {Phys. Rev. D}\ }\textbf {\bibinfo {volume} {94}},\ \bibinfo
  {pages} {034003} (\bibinfo {year} {2016})}\BibitemShut {NoStop}%
\bibitem [{\citenamefont {Brodsky}\ \emph {et~al.}(2002)\citenamefont
  {Brodsky}, \citenamefont {Hwang},\ and\ \citenamefont
  {Schmidt}}]{Brodsky:2002cx}%
  \BibitemOpen
  \bibfield  {author} {\bibinfo {author} {\bibfnamefont {S.~J.}\ \bibnamefont
  {Brodsky}}, \bibinfo {author} {\bibfnamefont {D.~S.}\ \bibnamefont {Hwang}},
  \ and\ \bibinfo {author} {\bibfnamefont {I.}~\bibnamefont {Schmidt}},\ }\href
  {\doibase 10.1016/S0370-2693(02)01320-5} {\bibfield  {journal} {\bibinfo
  {journal} {Phys. Lett. B}\ }\textbf {\bibinfo {volume} {530}},\ \bibinfo
  {pages} {99} (\bibinfo {year} {2002})}\BibitemShut {NoStop}%
\bibitem [{\citenamefont {Collins}(2002)}]{Collins:2002kn}%
  \BibitemOpen
  \bibfield  {author} {\bibinfo {author} {\bibfnamefont {J.~C.}\ \bibnamefont
  {Collins}},\ }\href {\doibase 10.1016/S0370-2693(02)01819-1} {\bibfield
  {journal} {\bibinfo  {journal} {Phys. Lett. B}\ }\textbf {\bibinfo {volume}
  {536}},\ \bibinfo {pages} {43} (\bibinfo {year} {2002})}\BibitemShut
  {NoStop}%
\bibitem [{\citenamefont {Ji}\ and\ \citenamefont {Yuan}(2002)}]{Ji:2002aa}%
  \BibitemOpen
  \bibfield  {author} {\bibinfo {author} {\bibfnamefont {X.-d.}\ \bibnamefont
  {Ji}}\ and\ \bibinfo {author} {\bibfnamefont {F.}~\bibnamefont {Yuan}},\
  }\href {\doibase 10.1016/S0370-2693(02)02384-5} {\bibfield  {journal}
  {\bibinfo  {journal} {Phys. Lett. B}\ }\textbf {\bibinfo {volume} {543}},\
  \bibinfo {pages} {66} (\bibinfo {year} {2002})}\BibitemShut {NoStop}%
\bibitem [{\citenamefont {Belitsky}\ \emph {et~al.}(2003)\citenamefont
  {Belitsky}, \citenamefont {Ji},\ and\ \citenamefont
  {Yuan}}]{Belitsky:2002sm}%
  \BibitemOpen
  \bibfield  {author} {\bibinfo {author} {\bibfnamefont {A.~V.}\ \bibnamefont
  {Belitsky}}, \bibinfo {author} {\bibfnamefont {X.}~\bibnamefont {Ji}}, \ and\
  \bibinfo {author} {\bibfnamefont {F.}~\bibnamefont {Yuan}},\ }\href {\doibase
  10.1016/S0550-3213(03)00121-4} {\bibfield  {journal} {\bibinfo  {journal}
  {Nucl. Phys.}\ }\textbf {\bibinfo {volume} {B656}},\ \bibinfo {pages} {165}
  (\bibinfo {year} {2003})}\BibitemShut {NoStop}%
\bibitem [{\citenamefont {Airapetian}\ \emph {et~al.}(2009)\citenamefont
  {Airapetian} \emph {et~al.}}]{Airapetian:2009ae}%
  \BibitemOpen
  \bibfield  {author} {\bibinfo {author} {\bibfnamefont {A.}~\bibnamefont
  {Airapetian}} \emph {et~al.} (\bibinfo {collaboration} {HERMES
  Collaboration}),\ }\href {\doibase 10.1103/PhysRevLett.103.152002} {\bibfield
   {journal} {\bibinfo  {journal} {Phys. Rev. Lett.}\ }\textbf {\bibinfo
  {volume} {103}},\ \bibinfo {pages} {152002} (\bibinfo {year}
  {2009})}\BibitemShut {NoStop}%
\bibitem [{\citenamefont {Grosse~Perdekamp}\ and\ \citenamefont
  {Yuan}(2015)}]{Perdekamp:2015vwa}%
  \BibitemOpen
  \bibfield  {author} {\bibinfo {author} {\bibfnamefont {M.}~\bibnamefont
  {Grosse~Perdekamp}}\ and\ \bibinfo {author} {\bibfnamefont {F.}~\bibnamefont
  {Yuan}},\ }\href {\doibase 10.1146/annurev-nucl-102014-021948} {\bibfield
  {journal} {\bibinfo  {journal} {Ann. Rev. Nucl. Part. Sci.}\ }\textbf
  {\bibinfo {volume} {65}},\ \bibinfo {pages} {429} (\bibinfo {year}
  {2015})}\BibitemShut {NoStop}%
\bibitem [{Bel()}]{BelleDetector}%
  \BibitemOpen
  \href@noop {} {}\bibinfo {howpublished} {A.~Abashian {\it et al.} (Belle
  Collaboration), Nucl. Instrum. Methods Phys. Res. Sect. A {\bf 479}, 117
  (2002); also see detector section in J. Brodzicka {\it et al.}, Prog. Theor.
  Exp. Phys. {\bf 2012}, 04D001 (2012).}\BibitemShut {Stop}%
\bibitem [{KEK()}]{KEKB}%
  \BibitemOpen
  \href@noop {} {}\bibinfo {howpublished} {S.~Kurokawa and E.~Kikutani, Nucl.
  Instrum. Methods Phys. Res. Sect. A {\bf 499}, 1 (2003), and other papers
  included in this Volume; T. Abe {\it et al.}, Prog. Theor. Exp. Phys. {\bf
  2013}, 03A001 (2013) and following articles up to 03A011}\BibitemShut
  {NoStop}%
\bibitem [{\citenamefont {Sj$\ddot{\rm{o}}$strand}\ \emph
  {et~al.}(2001)\citenamefont {Sj$\ddot{\rm{o}}$strand}, \citenamefont {Eden},
  \citenamefont {Friberg}, \citenamefont {Lonnblad}, \citenamefont {Miu},
  \citenamefont {Mrenna},\ and\ \citenamefont {Norrbin}}]{Sjostrand:2000wi}%
  \BibitemOpen
  \bibfield  {author} {\bibinfo {author} {\bibfnamefont {T.}~\bibnamefont
  {Sj$\ddot{\rm{o}}$strand}}, \bibinfo {author} {\bibfnamefont
  {P.}~\bibnamefont {Eden}}, \bibinfo {author} {\bibfnamefont {C.}~\bibnamefont
  {Friberg}}, \bibinfo {author} {\bibfnamefont {L.}~\bibnamefont {Lonnblad}},
  \bibinfo {author} {\bibfnamefont {G.}~\bibnamefont {Miu}}, \bibinfo {author}
  {\bibfnamefont {S.}~\bibnamefont {Mrenna}}, \ and\ \bibinfo {author}
  {\bibfnamefont {E.}~\bibnamefont {Norrbin}},\ }\href {\doibase
  10.1016/S0010-4655(00)00236-8} {\bibfield  {journal} {\bibinfo  {journal}
  {Comput. Phys. Commun.}\ }\textbf {\bibinfo {volume} {135}},\ \bibinfo
  {pages} {238} (\bibinfo {year} {2001})}\BibitemShut {NoStop}%
\bibitem [{\citenamefont {Lange}(2001)}]{Lange:2001uf}%
  \BibitemOpen
  \bibfield  {author} {\bibinfo {author} {\bibfnamefont {D.}~\bibnamefont
  {Lange}},\ }\href {\doibase 10.1016/S0168-9002(01)00089-4} {\bibfield
  {journal} {\bibinfo  {journal} {Nucl. Instrum. Methods Phys. Res. Sect. A}\
  }\textbf {\bibinfo {volume} {462}},\ \bibinfo {pages} {152} (\bibinfo {year}
  {2001})}\BibitemShut {NoStop}%
\bibitem [{\citenamefont {Brun}\ \emph {et~al.}(1987)\citenamefont {Brun},
  \citenamefont {Bruyant}, \citenamefont {Maire}, \citenamefont {McPherson},\
  and\ \citenamefont {Zanarini}}]{Brun:1987ma}%
  \BibitemOpen
  \bibfield  {author} {\bibinfo {author} {\bibfnamefont {R.}~\bibnamefont
  {Brun}}, \bibinfo {author} {\bibfnamefont {F.}~\bibnamefont {Bruyant}},
  \bibinfo {author} {\bibfnamefont {M.}~\bibnamefont {Maire}}, \bibinfo
  {author} {\bibfnamefont {A.}~\bibnamefont {McPherson}}, \ and\ \bibinfo
  {author} {\bibfnamefont {P.}~\bibnamefont {Zanarini}},\ }\href@noop {}
  {\bibfield  {journal} {\bibinfo  {journal} {CERN-DD-EE-84-1}\ } (\bibinfo
  {year} {1987})}\BibitemShut {NoStop}%
\bibitem [{\citenamefont {Seidl}\ \emph {et~al.}(2008)\citenamefont {Seidl}
  \emph {et~al.}}]{Seidl:2008xc}%
  \BibitemOpen
  \bibfield  {author} {\bibinfo {author} {\bibfnamefont {R.}~\bibnamefont
  {Seidl}} \emph {et~al.} (\bibinfo {collaboration} {Belle Collaboration}),\
  }\href {\doibase 10.1103/PhysRevD.78.032011, 10.1103/PhysRevD.86.039905}
  {\bibfield  {journal} {\bibinfo  {journal} {Phys. Rev. D}\ }\textbf {\bibinfo
  {volume} {78}},\ \bibinfo {pages} {032011} (\bibinfo {year}
  {2008})}\BibitemShut {NoStop}%
\bibitem [{\citenamefont {Vossen}\ \emph {et~al.}(2011)\citenamefont {Vossen}
  \emph {et~al.}}]{Vossen:2011fk}%
  \BibitemOpen
  \bibfield  {author} {\bibinfo {author} {\bibfnamefont {A.}~\bibnamefont
  {Vossen}} \emph {et~al.} (\bibinfo {collaboration} {Belle Collaboration}),\
  }\href {\doibase 10.1103/PhysRevLett.107.072004} {\bibfield  {journal}
  {\bibinfo  {journal} {Phys. Rev. Lett.}\ }\textbf {\bibinfo {volume} {107}},\
  \bibinfo {pages} {072004} (\bibinfo {year} {2011})}\BibitemShut {NoStop}%
\bibitem [{\citenamefont {Patrignani}\ \emph {et~al.}(2016)\citenamefont
  {Patrignani} \emph {et~al.}}]{PDG}%
  \BibitemOpen
  \bibfield  {author} {\bibinfo {author} {\bibfnamefont {C.}~\bibnamefont
  {Patrignani}} \emph {et~al.} (\bibinfo {collaboration} {Particle Data
  Group}),\ }\href@noop {} {\bibfield  {journal} {\bibinfo  {journal} {Chin.
  Phys. C}\ }\textbf {\bibinfo {volume} {40}},\ \bibinfo {pages} {100001}
  (\bibinfo {year} {2016})}\BibitemShut {NoStop}%
\bibitem [{\citenamefont {H$\ddot{\rm o}$cker}\ and\ \citenamefont
  {Kartvelishvili}(1996)}]{svdunfold}%
  \BibitemOpen
  \bibfield  {author} {\bibinfo {author} {\bibfnamefont {A.}~\bibnamefont
  {H$\ddot{\rm o}$cker}}\ and\ \bibinfo {author} {\bibfnamefont
  {V.}~\bibnamefont {Kartvelishvili}},\ }\href {\doibase
  10.1016/0168-9002(95)01478-0} {\bibfield  {journal} {\bibinfo  {journal}
  {Nucl. Instrum. Meth. A}\ }\textbf {\bibinfo {volume} {372}},\ \bibinfo
  {pages} {469} (\bibinfo {year} {1996})}\BibitemShut {NoStop}%
\bibitem [{SM()}]{SM}%
  \BibitemOpen
  \href@noop {} {\ }\bibinfo {note} {See Supplemental Material at [URL will be
  inserted by publisher] for figures showing the composition of quark
  flavors.}\BibitemShut {Stop}%
\bibitem [{sta()}]{state}%
  \BibitemOpen
  \href@noop {} {\ }\bibinfo {note} {Note, here the discussions focus on
  $\Lambda$, the contributions of the various quark flavors for $\bar{\Lambda}$
  can be inferred considering charge conjugation.}\BibitemShut {Stop}%
\bibitem [{\citenamefont {Niiyama}\ \emph {et~al.}(2018)\citenamefont {Niiyama}
  \emph {et~al.}}]{Niiyama:2017wpp}%
  \BibitemOpen
  \bibfield  {author} {\bibinfo {author} {\bibfnamefont {M.}~\bibnamefont
  {Niiyama}} \emph {et~al.} (\bibinfo {collaboration} {Belle Collaboration}),\
  }\href {\doibase 10.1103/PhysRevD.97.072005} {\bibfield  {journal} {\bibinfo
  {journal} {Phys. Rev. D}\ }\textbf {\bibinfo {volume} {97}},\ \bibinfo
  {pages} {072005} (\bibinfo {year} {2018})}\BibitemShut {NoStop}%
\bibitem [{\citenamefont {Lundberg}\ \emph {et~al.}(1989)\citenamefont
  {Lundberg} \emph {et~al.}}]{Lundberg:1989hw}%
  \BibitemOpen
  \bibfield  {author} {\bibinfo {author} {\bibfnamefont {B.}~\bibnamefont
  {Lundberg}} \emph {et~al.},\ }\href {\doibase 10.1103/PhysRevD.40.3557}
  {\bibfield  {journal} {\bibinfo  {journal} {Phys. Rev. D}\ }\textbf {\bibinfo
  {volume} {40}},\ \bibinfo {pages} {3557} (\bibinfo {year}
  {1989})}\BibitemShut {NoStop}%
\bibitem [{\citenamefont {Ramberg}\ \emph {et~al.}(1994)\citenamefont {Ramberg}
  \emph {et~al.}}]{Ramberg:1994tk}%
  \BibitemOpen
  \bibfield  {author} {\bibinfo {author} {\bibfnamefont {E.~J.}\ \bibnamefont
  {Ramberg}} \emph {et~al.},\ }\href {\doibase 10.1016/0370-2693(94)91397-8}
  {\bibfield  {journal} {\bibinfo  {journal} {Phys. Lett. B}\ }\textbf
  {\bibinfo {volume} {338}},\ \bibinfo {pages} {403} (\bibinfo {year}
  {1994})}\BibitemShut {NoStop}%
\bibitem [{\citenamefont {Abt}\ \emph {et~al.}(2006)\citenamefont {Abt} \emph
  {et~al.}}]{Abt:2006da}%
  \BibitemOpen
  \bibfield  {author} {\bibinfo {author} {\bibfnamefont {I.}~\bibnamefont
  {Abt}} \emph {et~al.} (\bibinfo {collaboration} {HERA-B Collaboration}),\
  }\href {\doibase 10.1016/j.physletb.2006.05.040} {\bibfield  {journal}
  {\bibinfo  {journal} {Phys. Lett. B}\ }\textbf {\bibinfo {volume} {638}},\
  \bibinfo {pages} {415} (\bibinfo {year} {2006})}\BibitemShut {NoStop}%
\bibitem [{\citenamefont {Fanti}\ \emph {et~al.}(1999)\citenamefont {Fanti}
  \emph {et~al.}}]{Fanti:1998px}%
  \BibitemOpen
  \bibfield  {author} {\bibinfo {author} {\bibfnamefont {V.}~\bibnamefont
  {Fanti}} \emph {et~al.},\ }\href {\doibase 10.1007/s100520050337} {\bibfield
  {journal} {\bibinfo  {journal} {Eur. Phys. J. C}\ }\textbf {\bibinfo {volume}
  {6}},\ \bibinfo {pages} {265} (\bibinfo {year} {1999})}\BibitemShut {NoStop}%
\bibitem [{\citenamefont {Aad}\ \emph {et~al.}(2015)\citenamefont {Aad} \emph
  {et~al.}}]{ATLAS:2014ona}%
  \BibitemOpen
  \bibfield  {author} {\bibinfo {author} {\bibfnamefont {G.}~\bibnamefont
  {Aad}} \emph {et~al.} (\bibinfo {collaboration} {ATLAS Collaboration}),\
  }\href {\doibase 10.1103/PhysRevD.91.032004} {\bibfield  {journal} {\bibinfo
  {journal} {Phys. Rev. D}\ }\textbf {\bibinfo {volume} {91}},\ \bibinfo
  {pages} {032004} (\bibinfo {year} {2015})}\BibitemShut {NoStop}%
\end{thebibliography}%





\clearpage
\begin{widetext}
{\huge{
Supplement to the publication: Observation of Transverse $\Lambda/\bar{\Lambda}$ Hyperon Polarization in $e^+e^-$ Annihilation at Belle}}
\vspace{3cm}
\end{widetext}

This supplement provides more detailed information accompanying the Letter ``Observation of Transverse $\Lambda/\bar{\Lambda}$ Hyperon Polarization in $e^+e^-$ Annihilation at Belle".

In Fig.~\ref{fig:Lmass}, we show the invariant mass of $p$ and $\pi^-$, where a clear  $\Lambda$ signal can be seen.
Fig.~\ref{fig:costheta} and \ref{fig:svd_costhe} display the cos$\theta$ distributions and the svd-unfolded cos$\theta$ distributions, respectively, for two bins in the thrust frame: $0.4 < z_\Lambda < 0.5$; $0.2< p_{\rm t}$ (GeV$/c$) $<0.5$  and $ 0.4 < z_\Lambda < 0.5$; $0.5<p_{\rm t}$ (GeV$/c$) $<0.8$.
The numerical results of transverse polarizations of $\Lambda$'s observed in the thrust frame and with associated production in the hadron frame are listed in Table~\ref{table:pol_vszpt_sys} and Table~\ref{table:pol_zlzh_sys}, respectively. Table~\ref{table:pol_bkgunfolded} displays the transverse polarizations for measured inclusive $\Lambda$'s, and the unfolded results for prompt $\Lambda$'s in the $uds$ sample and $\Lambda$'s from $\Sigma^0$ decays in the $uds$ sample in the thrust frame.

We investigate the flavor of the (anti-)quark going into the same hemisphere with the $\Lambda$ using MC, which is generated by Pythia 6.2.
The composition of quark flavors is displayed in Fig.~\ref{fig:zpt_mc} for inclusive $\Lambda$'s and Fig.~\ref{fig:zlpi_mc} for $\Lambda$'s in the associated production with light hadrons ($\pi^{\pm}$, $K^{\pm}$).

\begin{figure*}[h]
\begin{center}
\includegraphics[width=0.6\textwidth]{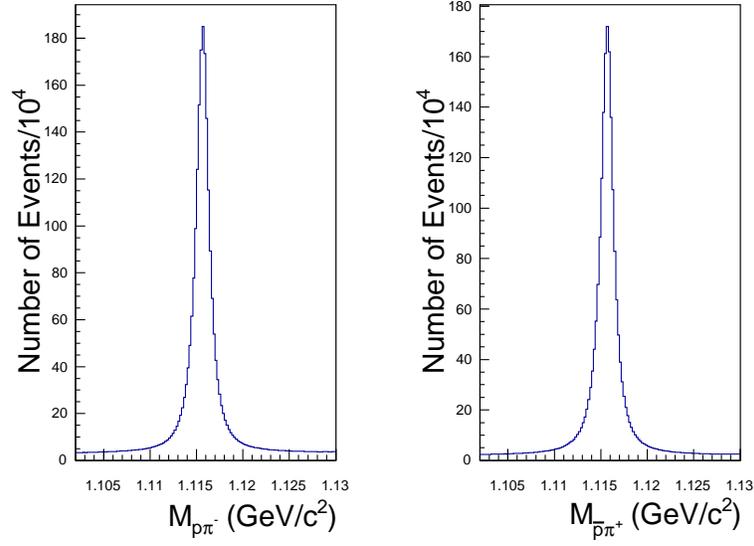}
\caption{Distributions of invariant mass of $p$ and $\pi^-$ as $\Lambda$ candidates (left) and invariant mass of $\bar{p}$ and $\pi^+$ as $\bar{\Lambda}$ candidates (right). 
\label{fig:Lmass}}
\end{center}
\end{figure*}

\begin{figure*}
\begin{center}
\includegraphics[width=0.6\textwidth]{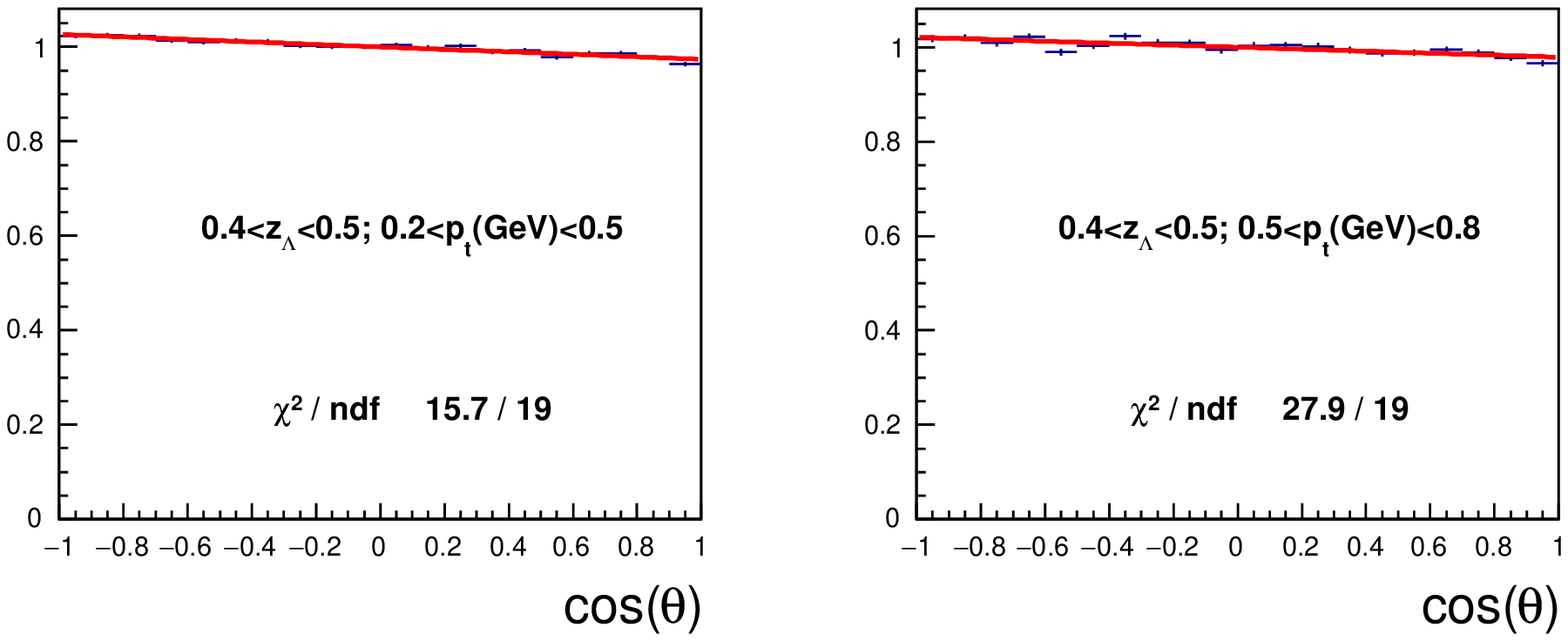}
\includegraphics[width=0.6\textwidth]{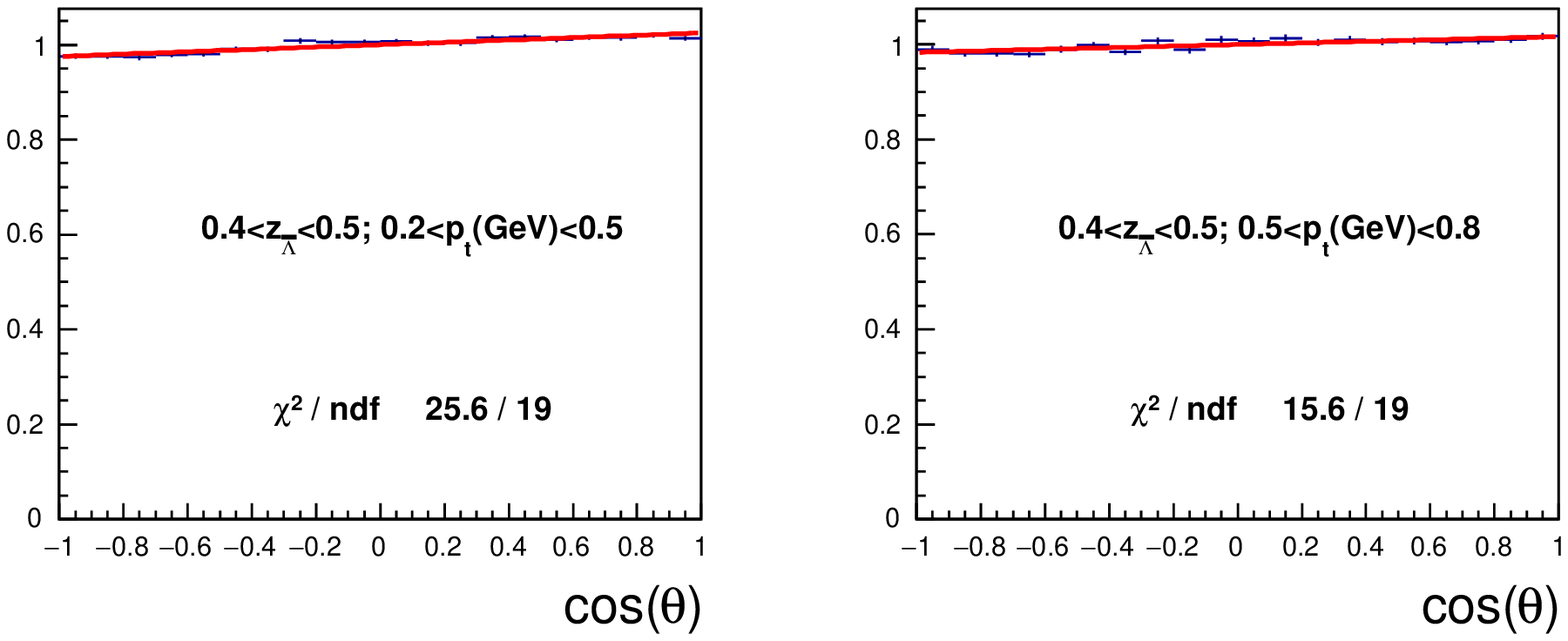}
\caption{The efficiency-corrected and normalized cos$\theta$ distributions in two specific [$z_\Lambda$, $p_{\rm t}$] bins. 
The top and bottom plots show the distributions for inclusive $\Lambda$ and $\bar{\Lambda}$, respectively, where the detector efficiencies are estimated using MC samples.} 
\label{fig:costheta}
\end{center}
\end{figure*}

\begin{figure*}
\begin{center}
\includegraphics[width=0.6\textwidth]{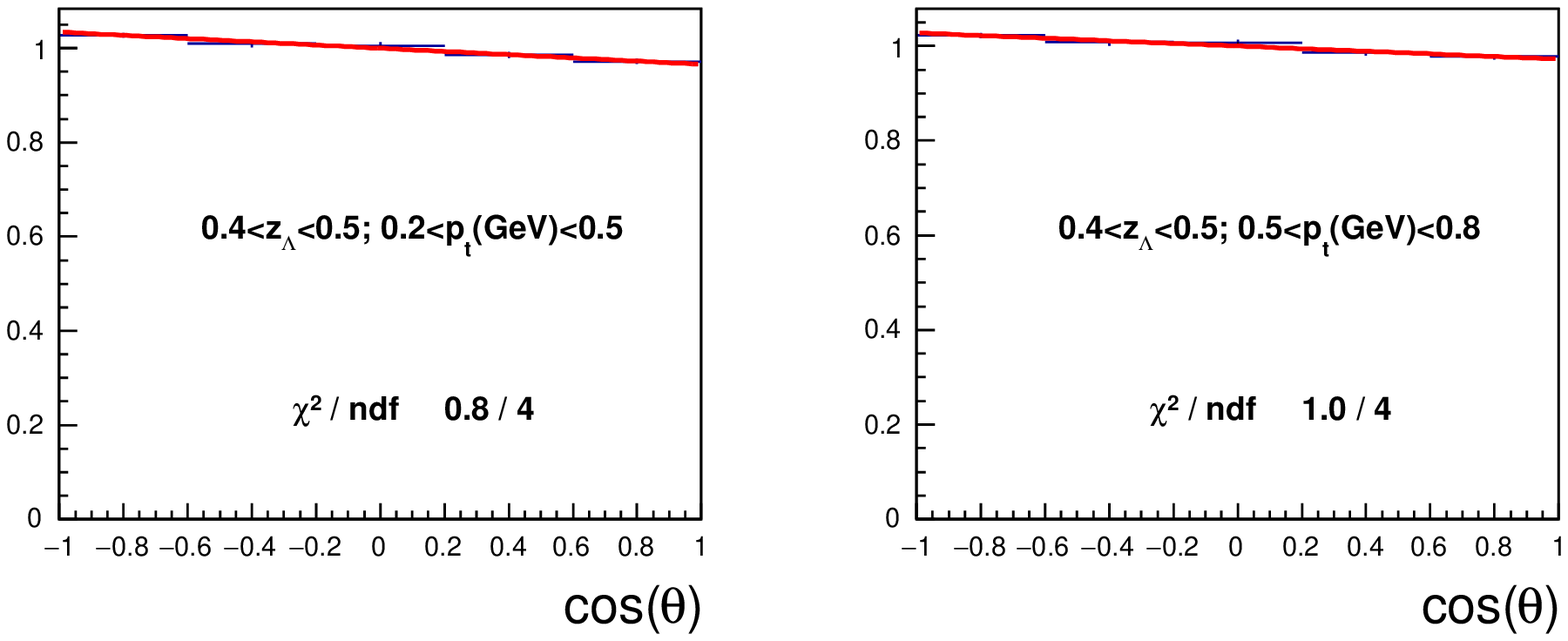}
\includegraphics[width=0.6\textwidth]{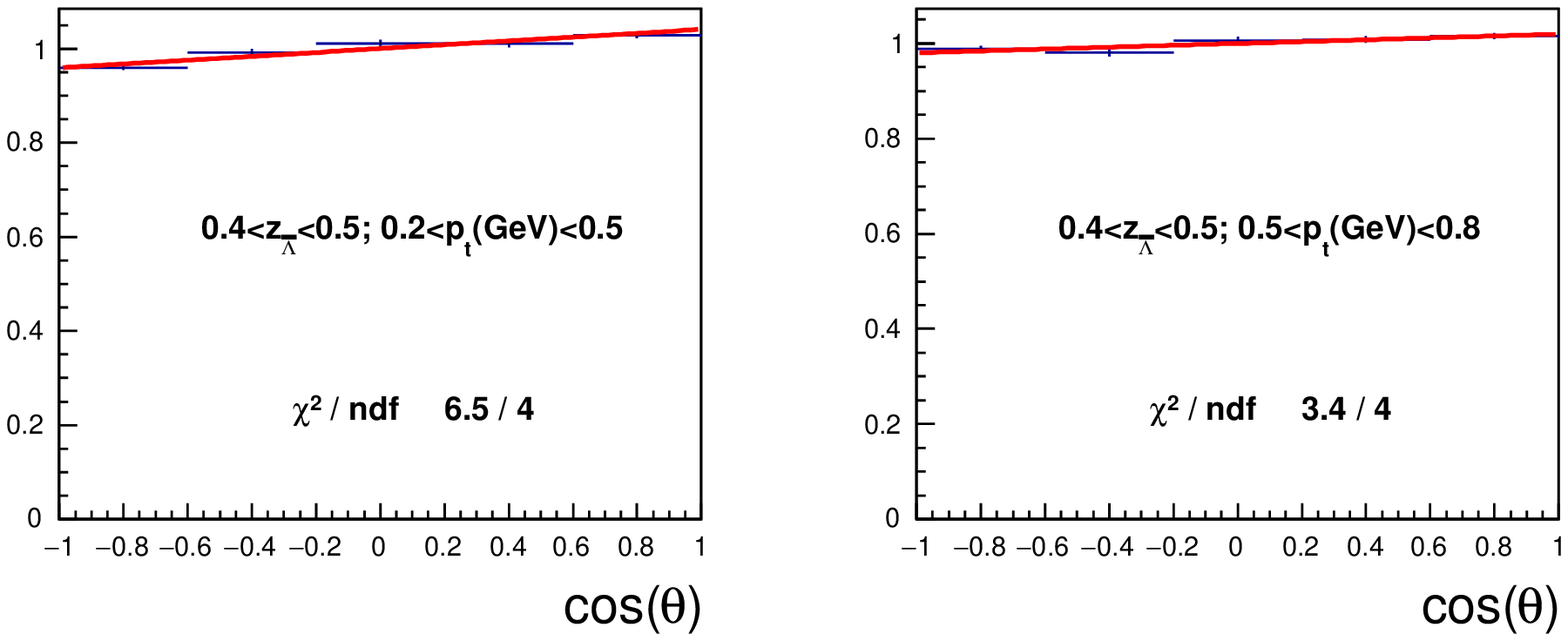}
\caption{The svd-unfolded and normalized cos$\theta$ distributions in two specific [$z_\Lambda$, $p_{\rm t}$] bins for inclusive $\Lambda$ (top) and $\bar{\Lambda}$ (bottom). }
\label{fig:svd_costhe}
\end{center}
\end{figure*}

\begin{table*}
\scriptsize
\begin{center}
\caption{Transverse polarizations, expressed in percent, observed in different [$z_{\Lambda}$, $p_{\rm t}$] bins for inclusive $\Lambda$ and $\bar{\Lambda}$ in the thrust frame. The first, second and third uncertainties are statistical, systematic and due to $\alpha$, respectively. The second and third columns list the averaged $z_{\Lambda}$ and $p_{\rm t}$.}
\label{table:pol_vszpt_sys}
\begin{tabular}{c|c|c|c|c}
\hline
 $z_{\Lambda}$, $p_{\rm t}$ (${\rm{GeV}}/c$)  & $<{z}_{\Lambda}>$ & $<{p}_{\rm t}>$  (${\rm{GeV}}/c$) & $\Lambda$ (\%)  &   $\bar{\Lambda}$ (\%)   \\
\hline
~[0.2, 0.3], [0.0, 0.2] & 0.25 & 0.13 &  1.98 $\pm$ 0.65 $\pm$ 0.31 $\pm$ 0.04  & -0.09 $\pm$ 0.71 $\pm$ 0.26 $\pm$ 0.00 \\ 
~[0.2, 0.3], [0.2, 0.5] & 0.25 & 0.34 & -0.78 $\pm$ 0.24 $\pm$ 0.15 $\pm$ 0.02  & -0.69 $\pm$ 0.25 $\pm$ 0.18 $\pm$ 0.01 \\
~[0.2, 0.3], [0.5, 0.8] & 0.26 & 0.60 & -0.33 $\pm$ 0.32 $\pm$ 0.16 $\pm$ 0.01  & -0.34 $\pm$ 0.35 $\pm$ 0.21 $\pm$ 0.01 \\
~[0.2, 0.3], [0.8, 1.6] & 0.28 & 0.88 & -1.34 $\pm$ 0.88 $\pm$ 0.45 $\pm$ 0.03  & -0.02 $\pm$ 0.95 $\pm$ 0.56 $\pm$ 0.00 \\
~[0.3, 0.4], [0.0, 0.2] & 0.34 & 0.13 & -3.01 $\pm$ 1.33 $\pm$ 0.54 $\pm$ 0.06  & -2.29 $\pm$ 1.42 $\pm$ 0.61 $\pm$ 0.05 \\
~[0.3, 0.4], [0.2, 0.5] & 0.35 & 0.34 & -2.51 $\pm$ 0.35 $\pm$ 0.14 $\pm$ 0.05  & -2.29 $\pm$ 0.37 $\pm$ 0.17 $\pm$ 0.05 \\
~[0.3, 0.4], [0.5, 0.8] & 0.35 & 0.62 &  0.04 $\pm$ 0.42 $\pm$ 0.18 $\pm$ 0.00  & -0.14 $\pm$ 0.45 $\pm$ 0.20 $\pm$ 0.00 \\
~[0.3, 0.4], [0.8, 1.6] & 0.35 & 0.96 & -1.17 $\pm$ 0.63 $\pm$ 0.30 $\pm$ 0.02  &  0.38 $\pm$ 0.66 $\pm$ 0.33 $\pm$ 0.01 \\
~[0.4, 0.5], [0.0, 0.2] & 0.44 & 0.13 & -6.83 $\pm$ 2.55 $\pm$ 1.25 $\pm$ 0.14  & -3.08 $\pm$ 2.68 $\pm$ 1.52 $\pm$ 0.06 \\
~[0.4, 0.5], [0.2, 0.5] & 0.44 & 0.35 & -5.34 $\pm$ 0.66 $\pm$ 0.27 $\pm$ 0.11  & -6.37 $\pm$ 0.70 $\pm$ 0.29 $\pm$ 0.13 \\
~[0.4, 0.5], [0.5, 0.8] & 0.45 & 0.63 & -4.31 $\pm$ 0.75 $\pm$ 0.31 $\pm$ 0.09  & -3.03 $\pm$ 0.79 $\pm$ 0.34 $\pm$ 0.06 \\
~[0.4, 0.5], [0.8, 1.6] & 0.45 & 0.99 & -2.28 $\pm$ 0.93 $\pm$ 0.44 $\pm$ 0.05  & -1.51 $\pm$ 0.95 $\pm$ 0.42 $\pm$ 0.03 \\
~[0.5, 0.9], [0.0, 0.2] & 0.59 & 0.13 & -3.98 $\pm$ 4.11 $\pm$ 2.41 $\pm$ 0.08  &  9.12 $\pm$ 4.22 $\pm$ 3.93 $\pm$ 0.18 \\
~[0.5, 0.9], [0.2, 0.5] & 0.59 & 0.35 & -4.66 $\pm$ 1.19 $\pm$ 0.87 $\pm$ 0.09  & -7.43 $\pm$ 1.23 $\pm$ 0.51 $\pm$ 0.15 \\
~[0.5, 0.9], [0.5, 0.8] & 0.59 & 0.63 & -8.76 $\pm$ 1.18 $\pm$ 0.60 $\pm$ 0.18  & -7.43 $\pm$ 1.21 $\pm$ 0.54 $\pm$ 0.15 \\
~[0.5, 0.9], [0.8, 1.6] & 0.60 & 1.04 & -6.70 $\pm$ 1.37 $\pm$ 0.56 $\pm$ 0.14  & -7.41 $\pm$ 1.38 $\pm$ 0.73 $\pm$ 0.15 \\
\hline
\end{tabular}
\end{center}
\end{table*}

\begin{table*}
\scriptsize
\begin{center}
\caption{Transverse polarizations of $\Lambda (\bar{\Lambda})$ in percent observed in different [$z_{\Lambda(\bar{\Lambda})}$, $z_{h}$] bins for processes $e^+e^-\to \Lambda (\bar{\Lambda})   h^{\pm}  X$ ($h = \pi, K$) in the hadron frame. The first, second and third uncertainties are statistical, systematic and due to $\alpha$, respectively.}
\label{table:pol_zlzh_sys}
\begin{tabular}{c|c|c|c|c}
\hline
$z_{\Lambda}$, $z_{h}$  &  $\Lambda  \pi^+ X$ (\%) &  $\Lambda  \pi^- X$ (\%) &  $\Lambda   K^+ X$  (\%)  &   $\Lambda   K^-  X$  (\%)    \\
\hline
~[0.2, 0.3], [0.2, 0.3] &  -3.90 $\pm$ 0.40 $\pm$ 0.01 $\pm$ 0.08  &  0.49 $\pm$ 0.38 $\pm$ 0.26 $\pm$ 0.01 &  -2.83 $\pm$ 0.50 $\pm$ 0.12 $\pm$ 0.06 &  -1.41 $\pm$ 0.61 $\pm$ 0.17 $\pm$ 0.03\\ 
~[0.2, 0.3], [0.3, 0.4] &  -4.74 $\pm$ 0.60 $\pm$ 0.18 $\pm$ 0.10  &  1.68 $\pm$ 0.58 $\pm$ 0.14 $\pm$ 0.03 &  -4.30 $\pm$ 0.76 $\pm$ 0.04 $\pm$ 0.09 &  -0.64 $\pm$ 0.88 $\pm$ 0.11 $\pm$ 0.01\\
~[0.2, 0.3], [0.4, 0.5] &  -7.08 $\pm$ 0.86 $\pm$ 0.13 $\pm$ 0.14  &  2.33 $\pm$ 0.85 $\pm$ 0.20 $\pm$ 0.05 &  -4.71 $\pm$ 1.07 $\pm$ 0.16 $\pm$ 0.10 &   2.70 $\pm$ 1.20 $\pm$ 0.43 $\pm$ 0.05\\
~[0.2, 0.3], [0.5, 0.9] &  -7.57 $\pm$ 0.99 $\pm$ 0.03 $\pm$ 0.15  &  3.72 $\pm$ 1.03 $\pm$ 0.31 $\pm$ 0.08 &  -12.02$\pm$ 1.10 $\pm$ 0.35 $\pm$ 0.24 &   4.27 $\pm$ 1.30 $\pm$ 0.45 $\pm$ 0.09\\
~[0.3, 0.4], [0.2, 0.3] &  -2.43 $\pm$ 0.39 $\pm$ 0.01 $\pm$ 0.05  & -0.20 $\pm$ 0.34 $\pm$ 0.07 $\pm$ 0.00 &  -0.11 $\pm$ 0.45 $\pm$ 0.07 $\pm$ 0.00 &  -1.60 $\pm$ 0.60 $\pm$ 0.09 $\pm$ 0.03\\
~[0.3, 0.4], [0.3, 0.4] &  -3.90 $\pm$ 0.60 $\pm$ 0.05 $\pm$ 0.08  & -0.32 $\pm$ 0.53 $\pm$ 0.04 $\pm$ 0.01 &  -4.08 $\pm$ 0.66 $\pm$ 0.08 $\pm$ 0.08 &  -0.38 $\pm$ 0.91 $\pm$ 0.04 $\pm$ 0.01\\
~[0.3, 0.4], [0.4, 0.5] &  -4.95 $\pm$ 0.90 $\pm$ 0.07 $\pm$ 0.10  & -0.53 $\pm$ 0.79 $\pm$ 0.03 $\pm$ 0.01 &  -5.11 $\pm$ 0.94 $\pm$ 0.03 $\pm$ 0.10 &  -1.07 $\pm$ 1.30 $\pm$ 0.24 $\pm$ 0.02\\
~[0.3, 0.4], [0.5, 0.9] &  -7.33 $\pm$ 1.07 $\pm$ 0.09 $\pm$ 0.15  &  1.75 $\pm$ 0.97 $\pm$ 0.10 $\pm$ 0.04 &  -8.80 $\pm$ 1.04 $\pm$ 0.11 $\pm$ 0.18 &   2.30 $\pm$ 1.43 $\pm$ 0.06 $\pm$ 0.05\\
~[0.4, 0.5], [0.2, 0.3] &  -2.89 $\pm$ 0.51 $\pm$ 0.01 $\pm$ 0.06  & -1.13 $\pm$ 0.45 $\pm$ 0.02 $\pm$ 0.02 &  -2.94 $\pm$ 0.58 $\pm$ 0.06 $\pm$ 0.06 &  -1.20 $\pm$ 0.81 $\pm$ 0.05 $\pm$ 0.02\\
~[0.4, 0.5], [0.3, 0.4] &  -4.38 $\pm$ 0.80 $\pm$ 0.08 $\pm$ 0.09  & -2.31 $\pm$ 0.69 $\pm$ 0.06 $\pm$ 0.05 &  -4.25 $\pm$ 0.82 $\pm$ 0.10 $\pm$ 0.09 &  -2.59 $\pm$ 1.25 $\pm$ 0.03 $\pm$ 0.05\\
~[0.4, 0.5], [0.4, 0.5] &  -5.81 $\pm$ 1.22 $\pm$ 0.12 $\pm$ 0.12  & -2.90 $\pm$ 1.03 $\pm$ 0.07 $\pm$ 0.06 &  -3.88 $\pm$ 1.13 $\pm$ 0.06 $\pm$ 0.08 &  -2.41 $\pm$ 1.78 $\pm$ 0.29 $\pm$ 0.05\\
~[0.4, 0.5], [0.5, 0.9] &  -6.51 $\pm$ 1.44 $\pm$ 0.16 $\pm$ 0.13  &  0.41 $\pm$ 1.28 $\pm$ 0.03 $\pm$ 0.01 &  -9.84 $\pm$ 1.21 $\pm$ 0.13 $\pm$ 0.20 &   1.70 $\pm$ 1.96 $\pm$ 0.06 $\pm$ 0.03\\
~[0.5, 0.9], [0.2, 0.3] &  -4.03 $\pm$ 0.65 $\pm$ 0.04 $\pm$ 0.08  & -2.47 $\pm$ 0.59 $\pm$ 0.00 $\pm$ 0.05 &  -2.41 $\pm$ 0.74 $\pm$ 0.05 $\pm$ 0.05 &  -2.52 $\pm$ 1.08 $\pm$ 0.03 $\pm$ 0.05\\
~[0.5, 0.9], [0.3, 0.4] &  -4.52 $\pm$ 1.01 $\pm$ 0.09 $\pm$ 0.09  & -2.42 $\pm$ 0.90 $\pm$ 0.09 $\pm$ 0.05 &  -5.24 $\pm$ 1.02 $\pm$ 0.02 $\pm$ 0.11 &  -1.46 $\pm$ 1.64 $\pm$ 0.09 $\pm$ 0.03\\
~[0.5, 0.9], [0.4, 0.5] &  -2.48 $\pm$ 1.53 $\pm$ 0.13 $\pm$ 0.05  & -2.31 $\pm$ 1.33 $\pm$ 0.17 $\pm$ 0.05 &  -4.77 $\pm$ 1.36 $\pm$ 0.14 $\pm$ 0.10 &  -1.95 $\pm$ 2.31 $\pm$ 0.29 $\pm$ 0.04\\
~[0.5, 0.9], [0.5, 0.9] &  -4.94 $\pm$ 1.79 $\pm$ 0.08 $\pm$ 0.10  & -4.15 $\pm$ 1.60 $\pm$ 0.21 $\pm$ 0.08 &  -9.57 $\pm$ 1.35 $\pm$ 0.19 $\pm$ 0.19 &   0.29 $\pm$ 2.56 $\pm$ 0.53 $\pm$ 0.01\\
\hline
$z_{\bar{\Lambda}}$, $z_{h}$ & $\bar{\Lambda}  \pi^+ X $  (\%)   &  $\bar{\Lambda}   \pi^- X $  (\%) &  $\bar{\Lambda}    K^+  X $  (\%) &   $\bar{\Lambda}   K^- X$  (\%)    \\
 \hline
~[0.2, 0.3], [0.2, 0.3] &   1.57 $\pm$ 0.39 $\pm$ 0.20 $\pm$ 0.03 & -3.59 $\pm$ 0.40 $\pm$ 0.04 $\pm$ 0.07 & -1.30 $\pm$ 0.63 $\pm$ 0.02 $\pm$ 0.03 & -2.36 $\pm$ 0.52 $\pm$ 0.02 $\pm$ 0.05\\ 
~[0.2, 0.3], [0.3, 0.4] &   1.27 $\pm$ 0.59 $\pm$ 0.08 $\pm$ 0.03 & -4.86 $\pm$ 0.61 $\pm$ 0.04 $\pm$ 0.10 & -0.07 $\pm$ 0.91 $\pm$ 0.11 $\pm$ 0.00 & -4.53 $\pm$ 0.77 $\pm$ 0.02 $\pm$ 0.09\\
~[0.2, 0.3], [0.4, 0.5] &   2.53 $\pm$ 0.86 $\pm$ 0.19 $\pm$ 0.05 & -7.37 $\pm$ 0.88 $\pm$ 0.01 $\pm$ 0.15 &  1.15 $\pm$ 1.22 $\pm$ 0.29 $\pm$ 0.02 & -5.66 $\pm$ 1.08 $\pm$ 0.07 $\pm$ 0.11\\
~[0.2, 0.3], [0.5, 0.9] &   3.10 $\pm$ 1.04 $\pm$ 0.34 $\pm$ 0.06 & -10.47$\pm$ 1.02 $\pm$ 0.24 $\pm$ 0.21 &  6.01 $\pm$ 1.32 $\pm$ 0.23 $\pm$ 0.12 & -10.95$\pm$ 1.14 $\pm$ 0.09 $\pm$ 0.22\\
~[0.3, 0.4], [0.2, 0.3] &  -0.23 $\pm$ 0.35 $\pm$ 0.00 $\pm$ 0.00 & -2.49 $\pm$ 0.39 $\pm$ 0.03 $\pm$ 0.05 & -0.23 $\pm$ 0.62 $\pm$ 0.02 $\pm$ 0.00 & -0.89 $\pm$ 0.46 $\pm$ 0.00 $\pm$ 0.02\\
~[0.3, 0.4], [0.3, 0.4] &  -1.00 $\pm$ 0.54 $\pm$ 0.04 $\pm$ 0.02 & -3.11 $\pm$ 0.62 $\pm$ 0.04 $\pm$ 0.06 & -0.32 $\pm$ 0.94 $\pm$ 0.07 $\pm$ 0.01 & -2.83 $\pm$ 0.67 $\pm$ 0.06 $\pm$ 0.06\\
~[0.3, 0.4], [0.4, 0.5] &  -0.35 $\pm$ 0.81 $\pm$ 0.02 $\pm$ 0.01 & -5.36 $\pm$ 0.93 $\pm$ 0.11 $\pm$ 0.11 &  0.29 $\pm$ 1.34 $\pm$ 0.05 $\pm$ 0.01 & -4.80 $\pm$ 0.95 $\pm$ 0.09 $\pm$ 0.10\\
~[0.3, 0.4], [0.5, 0.9] &   0.16 $\pm$ 1.01 $\pm$ 0.03 $\pm$ 0.00 & -6.25 $\pm$ 1.10 $\pm$ 0.18 $\pm$ 0.13 & -0.94 $\pm$ 1.48 $\pm$ 0.04 $\pm$ 0.02 & -9.58 $\pm$ 1.07 $\pm$ 0.23 $\pm$ 0.19\\
~[0.4, 0.5], [0.2, 0.3] &  -1.79 $\pm$ 0.46 $\pm$ 0.03 $\pm$ 0.04 & -2.48 $\pm$ 0.52 $\pm$ 0.02 $\pm$ 0.05 & -1.50 $\pm$ 0.83 $\pm$ 0.07 $\pm$ 0.03 & -1.30 $\pm$ 0.59 $\pm$ 0.08 $\pm$ 0.03\\
~[0.4, 0.5], [0.3, 0.4] &  -1.99 $\pm$ 0.71 $\pm$ 0.05 $\pm$ 0.04 & -4.32 $\pm$ 0.82 $\pm$ 0.10 $\pm$ 0.09 & -0.18 $\pm$ 1.29 $\pm$ 0.04 $\pm$ 0.00 & -2.26 $\pm$ 0.84 $\pm$ 0.00 $\pm$ 0.05\\
~[0.4, 0.5], [0.4, 0.5] &  -2.18 $\pm$ 1.06 $\pm$ 0.02 $\pm$ 0.04 & -2.69 $\pm$ 1.27 $\pm$ 0.08 $\pm$ 0.05 & -4.65 $\pm$ 1.86 $\pm$ 0.22 $\pm$ 0.09 & -5.97 $\pm$ 1.15 $\pm$ 0.12 $\pm$ 0.12\\
~[0.4, 0.5], [0.5, 0.9] &  -0.68 $\pm$ 1.32 $\pm$ 0.04 $\pm$ 0.01 & -8.34 $\pm$ 1.53 $\pm$ 0.16 $\pm$ 0.17 & -1.41 $\pm$ 2.09 $\pm$ 0.30 $\pm$ 0.03 & -8.26 $\pm$ 1.25 $\pm$ 0.10 $\pm$ 0.17\\
~[0.5, 0.9], [0.2, 0.3] &  -2.45 $\pm$ 0.60 $\pm$ 0.03 $\pm$ 0.05 & -1.96 $\pm$ 0.67 $\pm$ 0.03 $\pm$ 0.04 & -0.84 $\pm$ 1.10 $\pm$ 0.03 $\pm$ 0.02 & -5.08 $\pm$ 0.75 $\pm$ 0.07 $\pm$ 0.10\\
~[0.5, 0.9], [0.3, 0.4] &  -1.76 $\pm$ 0.93 $\pm$ 0.06 $\pm$ 0.04 & -5.26 $\pm$ 1.06 $\pm$ 0.25 $\pm$ 0.11 & -2.13 $\pm$ 1.72 $\pm$ 0.12 $\pm$ 0.04 & -6.09 $\pm$ 1.04 $\pm$ 0.05 $\pm$ 0.12\\
~[0.5, 0.9], [0.4, 0.5] &  -2.05 $\pm$ 1.38 $\pm$ 0.05 $\pm$ 0.04 & -4.17 $\pm$ 1.61 $\pm$ 0.09 $\pm$ 0.08 & -5.77 $\pm$ 2.42 $\pm$ 0.30 $\pm$ 0.12 & -6.21 $\pm$ 1.39 $\pm$ 0.14 $\pm$ 0.13\\
~[0.5, 0.9], [0.5, 0.9] &  -1.73 $\pm$ 1.68 $\pm$ 0.04 $\pm$ 0.04 & -5.34 $\pm$ 1.92 $\pm$ 0.10 $\pm$ 0.11 & -5.10 $\pm$ 2.73 $\pm$ 0.35 $\pm$ 0.10 & -8.28 $\pm$ 1.39 $\pm$ 0.11 $\pm$ 0.17\\
\hline
\end{tabular}
\end{center}
\end{table*}

\begin{table*}
\scriptsize
\begin{center}
\caption{Transverse polarizations, expressed in percent, observed in different $z_{\Lambda}$ bins for directly measured inclusive $\Lambda$'s, and the unfolded values for prompt $\Lambda$'s in the $uds$ sample and $\Lambda$'s from $\Sigma^0$ decays in the $uds$ sample in the thrust frame. 
The first uncertainty is the sum of statistical and systematic uncertainties while the second reflects the uncertainty from $\alpha$.}
\label{table:pol_bkgunfolded}
\begin{tabular}{c|c|c|c|c|c|c}
\hline
 $z_{\Lambda}$ &  inclusive $\Lambda$  (\%)   & inclusive  $\bar{\Lambda}$  (\%)   & prompt  $\Lambda$  (\%) & prompt $\bar{\Lambda}$  (\%) & $\Lambda$  from $\Sigma^0$ decays  (\%) & $\bar\Lambda$ from $\bar{\Sigma}^0$ decays  (\%)  \\
 \hline
 ~[0.2, 0.3] & -0.64 $\pm$ 0.17 $\pm$ 0.01 & -0.27 $\pm$ 0.19 $\pm$ 0.01  & 2.79 $\pm$ 2.91 $\pm$ 0.06   &  0.08 $\pm$ 3.10 $\pm$ 0.00   &  1.31 $\pm$ 9.09 $\pm$ 0.03  &  9.56 $\pm$ 9.58 $\pm$ 0.19   \\ 
 ~[0.3, 0.4] & -1.33 $\pm$ 0.16 $\pm$ 0.03 & -1.27 $\pm$ 0.15 $\pm$ 0.03  & 1.88 $\pm$ 2.90 $\pm$ 0.04   &  5.84 $\pm$ 1.57 $\pm$ 0.12   &  -5.65 $\pm$ 9.91 $\pm$ 0.11 &  -5.87 $\pm$ 1.95 $\pm$ 0.12  \\
 ~[0.4, 0.5] & -4.35 $\pm$ 0.23 $\pm$ 0.09 & -3.75 $\pm$ 0.22 $\pm$ 0.08  & -8.74 $\pm$ 3.44 $\pm$ 0.18  &  -6.01 $\pm$ 3.53 $\pm$ 0.12  &  8.22 $\pm$ 10.04 $\pm$ 0.17 &  5.60 $\pm$ 10.30 $\pm$ 0.11  \\
 ~[0.5, 0.7] & -5.67 $\pm$ 0.33 $\pm$ 0.11 & -5.59 $\pm$ 0.34 $\pm$ 0.11  & -13.25 $\pm$ 4.08 $\pm$ 0.27 &  -10.69 $\pm$ 4.12 $\pm$ 0.22 &  4.26 $\pm$ 9.05 $\pm$ 0.09  &  0.06 $\pm$ 9.21 $\pm$ 0.00   \\
 ~[0.7, 0.9] & -6.98 $\pm$ 1.44 $\pm$ 0.14 & -7.47 $\pm$ 1.43 $\pm$ 0.15  & -9.29 $\pm$ 9.05 $\pm$ 0.19  &  -12.49 $\pm$ 2.32 $\pm$ 0.25 &  -4.03 $\pm$ 6.24 $\pm$ 0.08 &  -1.37 $\pm$ 18.13 $\pm$ 0.03 \\

\hline
\end{tabular}
\end{center}
\end{table*}

\begin{figure*}
\begin{center}
\includegraphics[width=0.72\textwidth]{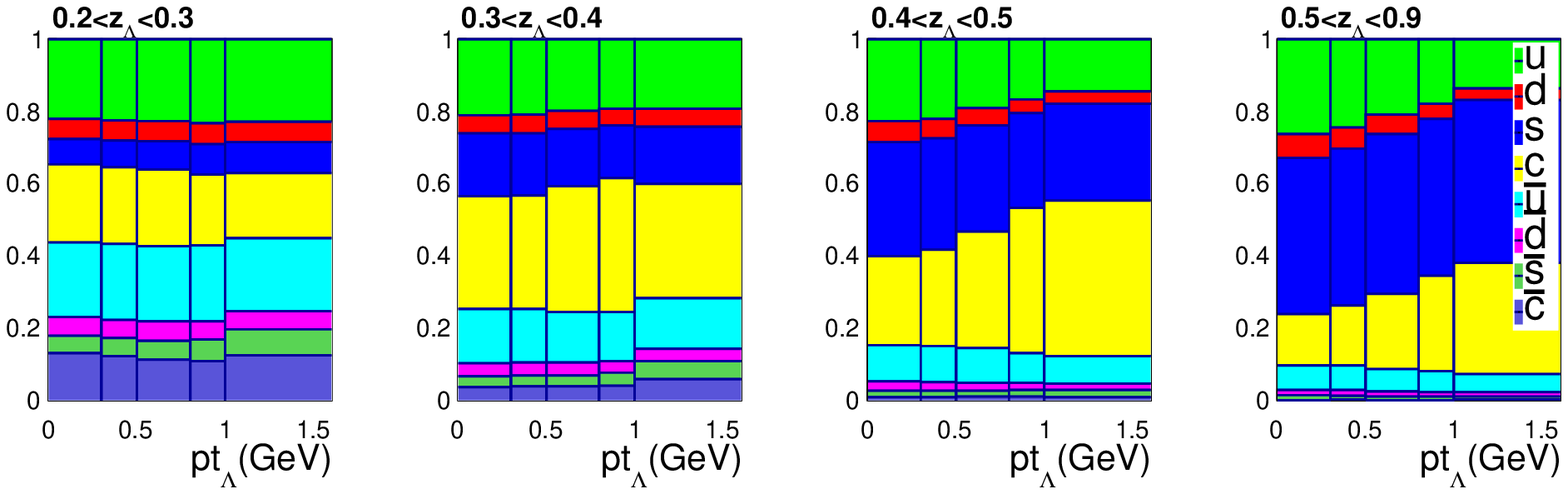}
\includegraphics[width=0.72\textwidth]{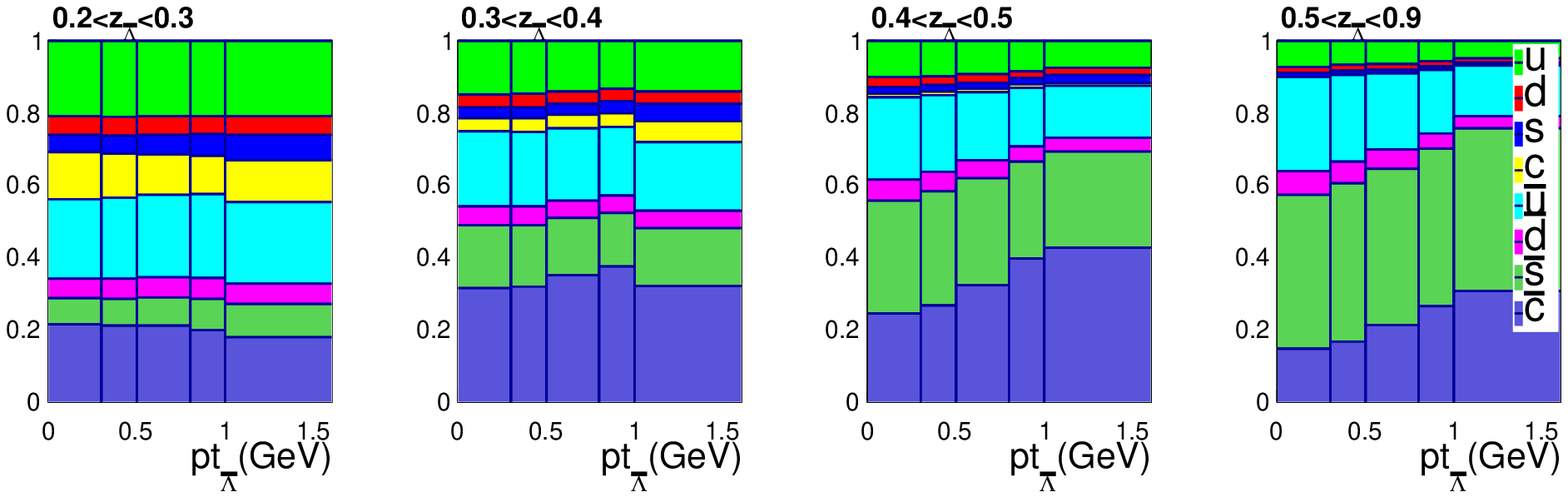}
\caption{The flavor of the quark going into the same hemisphere with the $\Lambda$ or $\bar{\Lambda}$ in the inclusive process $e^+e^- \rightarrow \Lambda  X$ (top) and $e^+e^- \rightarrow \bar{\Lambda}  X$ (bottom) in different [$z_{\Lambda}$, $p_t$] regions.  The Y axis shows the fractions from different quark flavors in a stacked style. } 
\label{fig:zpt_mc}
\end{center}
\end{figure*}

\begin{figure*}
\begin{center}
\includegraphics[width=0.72\textwidth]{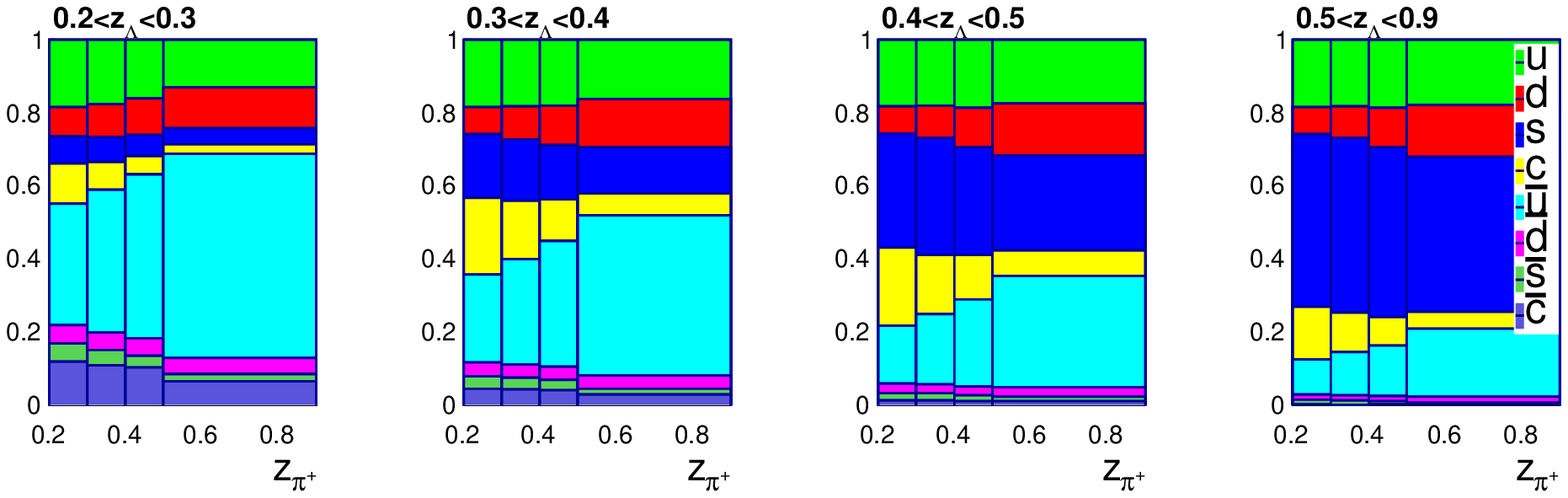}
\includegraphics[width=0.72\textwidth]{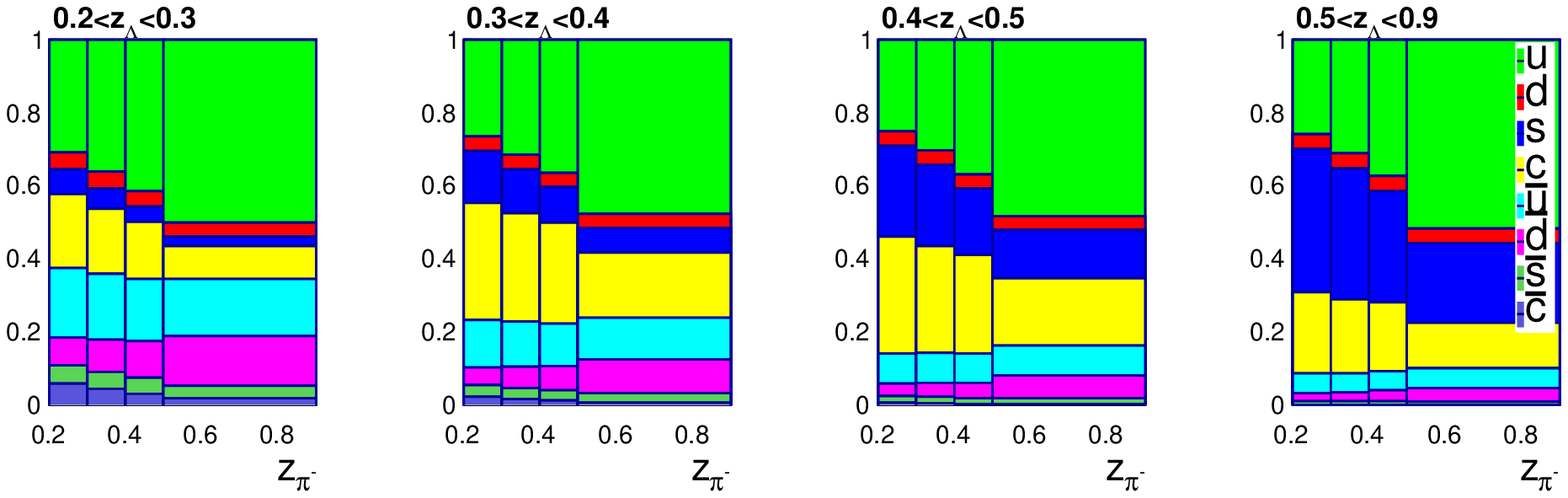}
\includegraphics[width=0.72\textwidth]{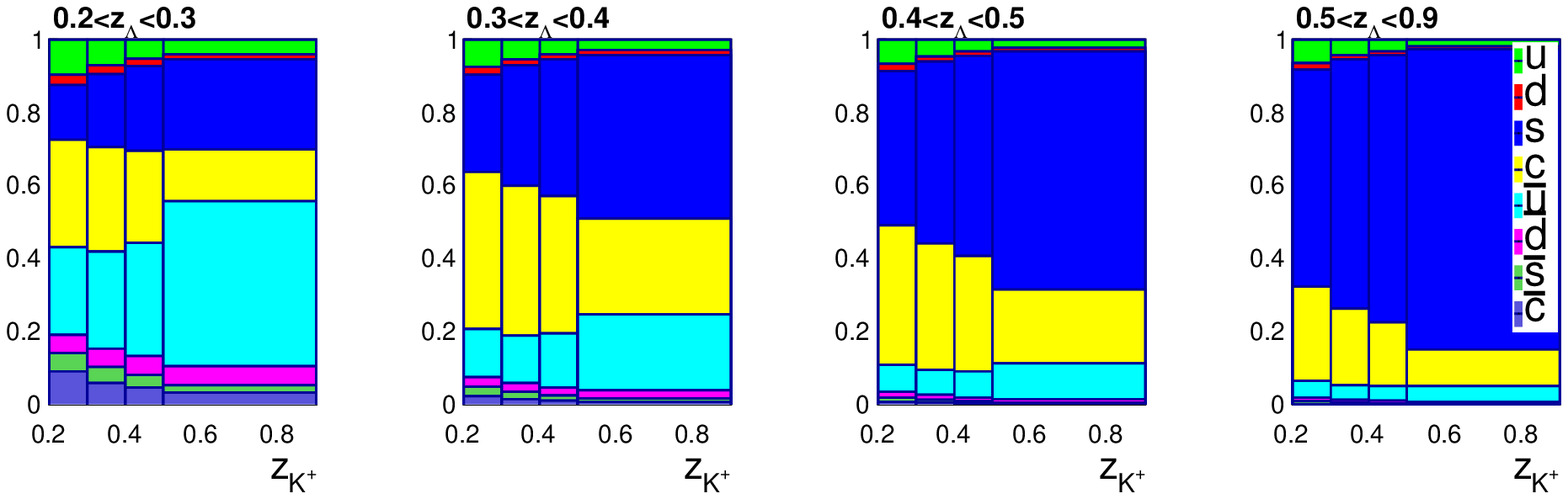}
\includegraphics[width=0.72\textwidth]{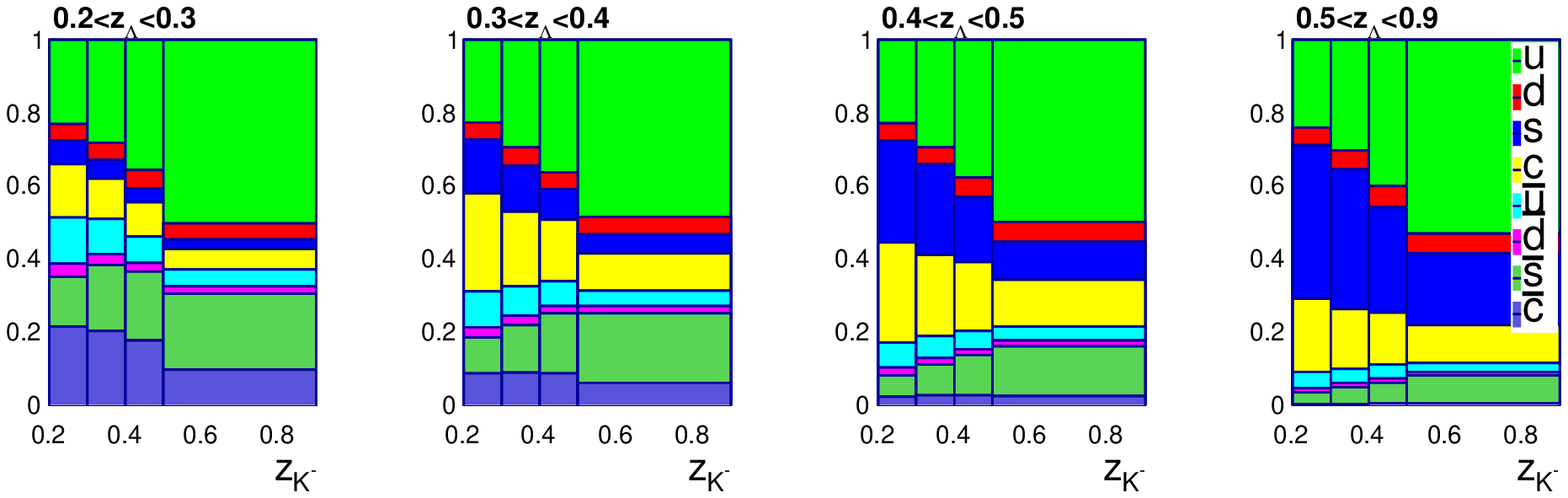}
\caption{The flavor of quark going into the same hemisphere with the $\Lambda$ in $e^+e^- \rightarrow \Lambda  h^{\pm}  X $ ($h=\pi, K$) in different [$z_\Lambda$, $z_{h^{\pm}}$] bins. 
The Y axis shows the fractions from different quark flavors in a stacked style. } 
\label{fig:zlpi_mc}
\end{center}
\end{figure*}


\end{document}